\documentclass[10pt,journal,compsoc]{IEEEtran}
\ifCLASSOPTIONcompsoc
  \usepackage[nocompress]{cite}
\else
  \usepackage{cite}
\fi
\ifCLASSINFOpdf
\else
\fi
\usepackage{etoolbox}
\usepackage{xspace}
\usepackage[table]{xcolor}
\usepackage{graphicx}
\usepackage[switch]{lineno}
\usepackage{fontawesome}
\usepackage{multirow}
\usepackage{url}
\usepackage{hyperref}

\begin{document}
\bstctlcite{IEEEexample:BSTcontrol}
\newtoggle{draft}
\toggletrue{draft}

\iftoggle{draft}
{
	\newcommand{\jon}[1]{{\leavevmode\color{red}{Jon: #1}\xspace}}
	\newcommand{\waqar}[1]{{\leavevmode\color{blue}{Waqar: #1}\xspace}}
	\newcommand{\rifat}[1]{{\leavevmode\color{violet}{Rifat: #1}\xspace}}
	\newcommand{\harsha}[1]{{\leavevmode\color{orange}{Harsha: #1}\xspace}}
	\newcommand{\arif}[1]{{\leavevmode\color{magenta}{Arif: #1}\xspace}}
	\newcommand{\gillian}[1]{{\leavevmode\color{brown}{Gillian: #1}\xspace}}
	\newcommand{\rashina}[1]{{\leavevmode\color{red}{Rashina: #1}\xspace}}
	\newcommand{\mojtaba}[1]{{\leavevmode\color{red}{Mojtaba: #1}\xspace}}
}
{
	\newcommand{\jon}[1]{}
	\newcommand{\waqar}[1]{}
	\newcommand{\rifat}[1]{}
	\newcommand{\harsha}[1]{}
	\newcommand{\arif}[1]{}
	\newcommand{\gillian}[1]{}
	\newcommand{\rashina}[1]{}
	\newcommand{\mojtaba}[1]{}
}

%
\title{How Can Human Values Be Addressed in Agile Methods? A Case Study on SAFe}
%
%
%
%


\author{Waqar~Hussain,~\IEEEmembership{}
        Mojtaba~Shahin,~\IEEEmembership{}
        Rashina~Hoda,\IEEEmembership{}
        Jon~Whittle,~\IEEEmembership{}
        Harsha~Perera,~\IEEEmembership{}
        Arif~Nurwidyantoro,~\IEEEmembership{}
        Rifat~Ara~Shams,~\IEEEmembership{}
        and~Gillian~Oliver~\IEEEmembership{}
\IEEEcompsocitemizethanks{\IEEEcompsocthanksitem 
Waqar Hussain, Mojtaba Shahin, Rashina Hoda, Harsha Perera, Arif Nurwidyantoro, Rifat Ara Shams and Gillian Oliver are with the Faculty of Information Technology, Monash University, Australia.
\protect\\
E-mail: \{Waqar.Hussain,~Mojtaba.Shahin,~Rashina.Hoda,~Harsha.Perera,~Arif.Nurwidyantoro,~Rifat.Shams,~Gillian.Oliver\}@monash.edu}\\
\IEEEcompsocitemizethanks{\IEEEcompsocthanksitem 
Jon Whittle is with CSIRO's Data61, Australia.
\protect\\
E-mail: Jon.Whittle@data61.csiro.au
}

\thanks{Manuscript submitted to IEEE Transactions on Software Engineering (2021)}
}
\markboth{}
{}
\IEEEtitleabstractindextext{
\begin{abstract}
Agile methods are predominantly focused on delivering business values. But can Agile methods be adapted to effectively address and deliver human values such as social justice, privacy, and sustainability in the software they produce? Human values are what an individual or a society considers important in life. Ignoring these human values in software can pose difficulties or risks for all stakeholders (e.g., user dissatisfaction, reputation damage, financial loss). 
To answer this question, we selected the Scaled Agile Framework (SAFe), one of the most commonly used Agile methods in the industry, and conducted a qualitative case study to identify possible intervention points within SAFe that are the most natural to address and integrate human values in software. 
We present five high-level empirically-justified sets of interventions in SAFe: \textit{artefacts}, \textit{roles}, \textit{ceremonies}, \textit{practices}, and \textit{culture}. We elaborate how some current Agile artefacts (e.g., user story), roles (e.g., product owner), ceremonies (e.g., stand-up meeting), and practices (e.g., business-facing testing) in SAFe can be modified to support the inclusion of human values in software. Further, our study suggests new and exclusive values-based artefacts (e.g., legislative requirement), ceremonies (e.g., values conversation), roles (e.g., values champion), and cultural practices (e.g., induction and hiring) to be introduced in SAFe for this purpose.
Guided by our findings, we argue that existing Agile methods can account for human values in software delivery with some evolutionary adaptations. 

\end{abstract}

}

\maketitle

\IEEEdisplaynontitleabstractindextext

%
\IEEEpeerreviewmaketitle

\section{Introduction}\label{sec:introduction}


Human values---such as justice, trust, privacy, honesty, sustainability, and the like are increasingly recognised as critical to software development~\cite{8917668}.
This realisation has in part come about as a result of high profile media cases where a failure to address human values has led to significant reputation damage, financial loss, and/or user dissatisfaction.
The Cambridge Analytica and Facebook scandal~\cite{neate_2018}, Google's Project Maven~\cite{shane_wakabayashi_2018}, and the use of social media platforms to spread propaganda during elections~\cite{bradshaw_howard}, are just a few cases where the software industry in general and large software firms in particular have failed to properly address a range of values such as trust, inclusion, fairness, empathy during software development. As a result, there are now countless examples of large-scale software systems that can be considered ethically dubious, morally suspicious, or simply not aligned to the value set of the user base~\cite{galhotra2017fairness}.

In addressing values in Software Engineering (SE), research and practice have thus far focused on a limited set of values such as security, privacy, and accessibility, largely ignoring the vast majority of human values such as social justice, cultural traditions, curiosity, and environmental sustainability~\cite{perera2020}. Perera et al.~\cite{perera2020} found that only 16\% of the publications between 2015 and 2018 in the leading SE venues were directly relevant to human values. Some of the reasons for the lack of attention to values range from a lack of diversity amongst the leaders in the technology industry, relentless drive to profit in an increasingly competitive market, to lack of legislation to regulate the use of new technologies~\cite{philbeck2018values}. 

Another key barrier to the development of more values-conscious software is the lack of guidelines in software development methodologies for addressing values. As a result, even when regulations are introduced---such as the EU's General Data Protection Regulation (GDPR)~\cite{EuGDPR} 
and various governmental frameworks on the ethical use of AI introduced internationally~\cite{raso2018artificial}---software practitioners struggle to implement those guidelines in practice. 
GDPR is overwhelmingly lengthy but, at the same time, not very precise. On the other hand, AI ethics frameworks
typically are relatively less drawn out but still very high level. While important in their own right, these two arguably `edge-of-spectrum' approaches illustrate the lack of actionable knowledge on how to introduce human values into existing software development processes and frameworks~\cite{sirur2018we}.



What is needed is practical and effective adaptations to everyday software development practices so that values can be explicitly and effectively addressed \textit{during} software development, and their absence is not retrospectively regretted or justified.

Agile software development methods are undeniably the most popular software development methodologies, driving small and large-scale development of software \cite{hoda2018rise}. Due to their focus on people and interactions~\cite{fowler2001agile}, agile methods are particularly amenable to incorporating human values. 
Of the various agile methods at scale such as Disciplined Agile Delivery (DAD), Large Scale Scrum (LeSS), Scaled Agile Framework (SAFe), and Spotify, SAFe \cite{SAFe2020} has gained prominence in recent years because of its focus on effectively applying core agile principles and a range of agile practices at scale within large organisations. 

Agile is a values-driven philosophy and its relationship with values is evident from the famous four value statements in Agile manifesto \cite{fowler2001agile}. The creators of Agile manifesto emphasised that: \textit{``the meteoric rise of interest in—and sometimes tremendous criticism of—Agile Methodologies is about the mushy stuff of values and culture''} \cite{fowler2001agile}. In essence, Agile methodology encompasses a number of human values such as trust, respect, collaboration, community, and delegation of authority or power etc. \cite{williams2003agile}. However, Agile implementation is generally restricted to the consideration and delivery of business or economic value \cite{alahyari2017study}. This has somewhat derailed Agile software development from the original purpose and (the values-rich) philosophy and thus needs to be linked back to it. Agile methodologies not only can accommodate human values but also encourage their adoption for development teams, as noted by Sutcliffe et al. \cite{sutcliffe2011experience}. Since SAFe utilises a combination of Agile methodologies such as Scrum and Kanban, it should, like other agile methods, adhere to the essence of Agile philosophy i.e. its values \cite{beck2001manifesto} \cite{boehm2002get}.




Previous research has focused on addressing values in a particular phase of the software development lifecycle (e.g., requirements elicitation \cite{thew2018value}, design \cite{van2015design}), identifying relations between Agile values and human values (e.g., \cite{fagerholm2014examining}), and detecting human values from app reviews (e.g., \cite{obie2021first}). This paper is comparatively unique as it aims to address human values in the software development process. We undertook a qualitative, exploratory case study \cite{yin2014case} to identify the intervention points within SAFe that appear to be more practical and natural to allow consideration of human values into the SAFe framework. In particular, our aim was to address the overarching research question: \textbf{\textit{where and how can human values be most effectively addressed in the Scaled Agile Framework?}}


Over a twelve-month period, our research 
team worked closely with the IT business unit of a major public organisation. This business unit (henceforth, the case company) develops a range of software products and services for use within the broader organisation. For the last three years, it has been undergoing an agile transformation, gradually introducing the SAFe framework. 
The case company currently uses the \textit{Essential} configuration of SAFe (see Section \ref{sec:AgileSAFe}) and combines it with practices from human centred design and design thinking.

The key contribution of this paper is empirically identifying five high-level intervention points, namely, \textit{artefacts}, \textit{roles}, \textit{ceremonies}, \textit{practices}, and \textit{culture}, to introduce human values in the SAFe framework. We specifically articulate the modifications required for each of these intervention points to ensure human values are considered as primary artefacts in software development practice. Moreover, we introduce a set of new values-based artefacts, roles, ceremonies, and cultural practices that can help development organisations transition from being mainly \textit{productivity} or \textit{business} value-driven towards having a \textit{human} values-oriented development mindset.

\textbf{The Paper Organisation:} Section \ref{sec:BackRelatedWork} provides the background for this study and summarises the related research. We describe our research method in Section \ref{sec:CaseStudyDesign}. Section \ref{sec:Findings} reports our findings which we discuss and reflect upon in Section \ref{sec:Discussion}. The possible threats to the validity of our study are discussed in Section \ref{sec:Threats}. Finally, Section \ref{sec:Conclusion} concludes our study with some research directions for the future.  
\section{Background and Related Works}\label{sec:BackRelatedWork}



\subsection{Human Values}\label{sec:ValueDefinition}
Values are ``deep-rooted personal criteria on which thoughts and actions are, often unconsciously, based and evaluated'' \cite{bouman2018measuring}. They are human perceptions of the importance of learning (something) and acting or behaving in given ways \cite{usher2017social}. Every human possesses a set of values \cite{rokeach1973nature} such as safety, sustainability, family, freedom, equality, etc., that may be rooted in language, religion, philosophies, and communal culture.

At the root of human values are three common yet basic human needs  --- \textit{`human survival as biological organisms'}, \textit{`requisites of coordinated social interaction'}, and \textit{`survival and welfare of groups'}~\cite{schwartz1992universals}. To some extent, most individuals endorse a similar set of values, but they differ in how they prioritise one value over another. For example, one might value environmental sustainability more than the gratification of personal desires and choose a more (personally perceived) responsible food choice (e.g., vegan) compared to other less socially responsible alternatives \cite{mussbacher2020there}. People often act in ways that lead to outcomes they value and avoid actions that may result in outcomes that are inconsistent with their values \cite{schunk2012social}. This pursuit of `values-alignment' arguably provides them a sense of fulfilment \cite{palacin2021human}.

\begin{figure}
    \centering
    \includegraphics[width=0.48\textwidth]{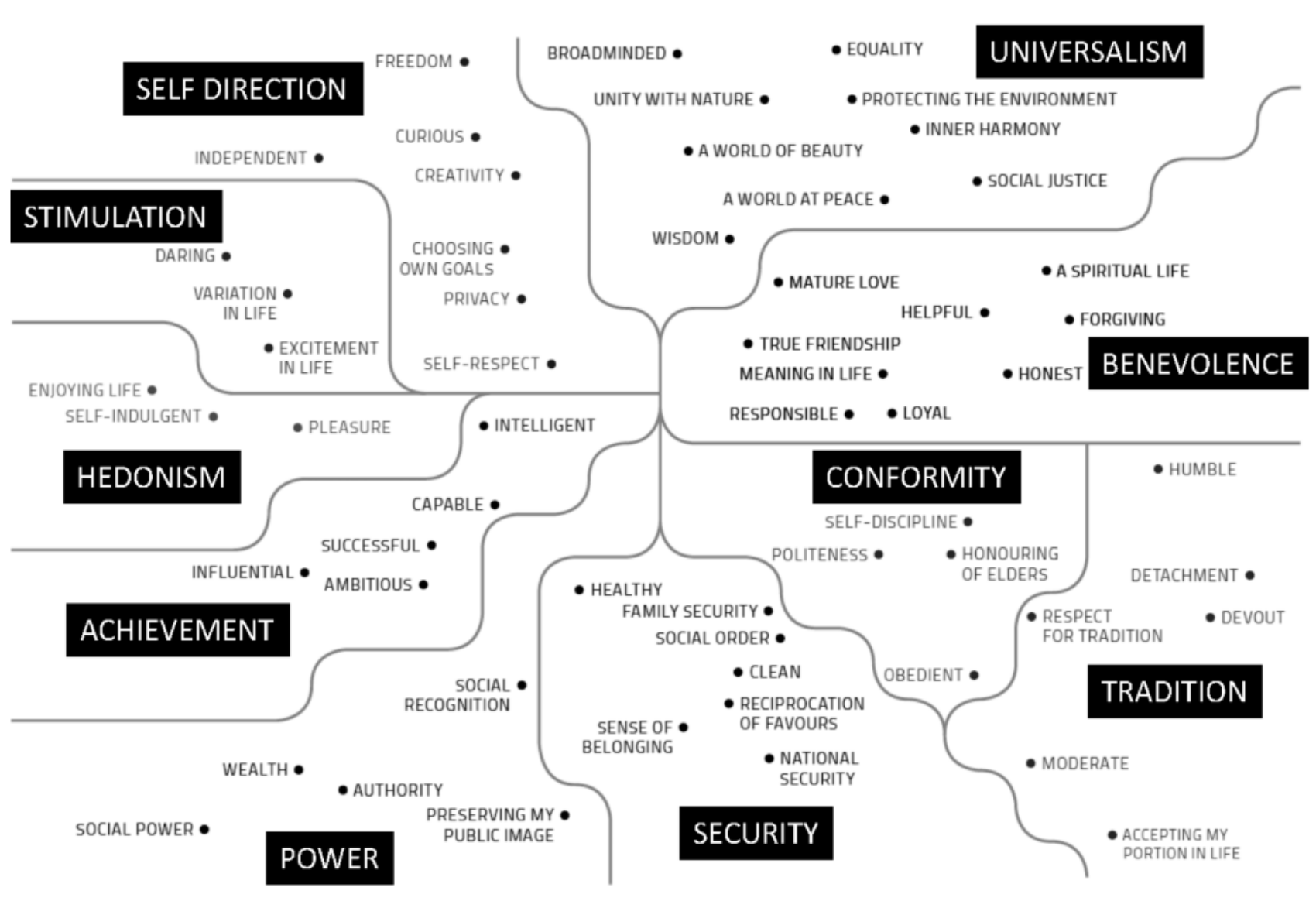}
    \caption{Schwartz values model \cite{schwartz1992universals} (adopted from \cite{CommonCauseHandbook}, \cite{ferrario2016values}). The boxes indicate the ten basic human values, and each of them is subdivided 
  into some finer-grained motivational goals or value items. 
    }
    \label{fig:SchwartzModel}
\end{figure}

The question that has perplexed many researchers is whether human values are shared universally. Schwartz's \textbf{\textit{Theory of Basic Human Values}}\cite{schwartz1992universals}, the most comprehensive and empirically validated theory of values to date, is the closest to answer the question. 
Using surveys from 68 cultures across the world \cite{whittle2019case}, the theory postulates ten (near-universal) basic human values such as Power, Achievement, Security, Universalism, Benevolence, etc. (the black boxes in Figure \ref{fig:SchwartzModel}). These values are distinguished by their corresponding motivational goals or value items, such as \textit{wealth}, \textit{successful}, \textit{family security}, etc. (56 in total, see Figure \ref{fig:SchwartzModel}). The theory allows visualisation of a broader set of values and their underlying motivational goals that do not necessarily have an ethical or moral import (e.g., power, hedonism, wealth, achievement, or pleasure etc.) but may concern stakeholders and inform technological design.  To explain the dynamic relationship among values, Schwartz explicated a circular structure that captures the conflicts and compatibility among these values. For example, value items adjacent to each other share a common motivation, such as \textit{broad-mindedness} and \textit{freedom}, while those in opposing directions are in tension, such as \textit{obedience} and \textit{freedom} ~\cite{schwartz1994there}. This circular structure enables the evaluation of value conflicts and trade-offs \cite{9218163}. Some values such as privacy, security, responsibility, and fairness are actively pursued in Software Engineering, Human-Computer Interactions, and Artificial Intelligence \cite{perera2020}.

The pursuit of imparting values in technology should not be surprising. In 1980, Langdon Winner, in his classic paper \cite{Winner1980}, `Do artefacts have politics?', highlighted that technology is not values-neutral. His main argument was that designers, through their choices and decisions can embed values such as power and authority into technological artefacts e.g., software design \cite{Winner1980}. From the HCI domain in the nineties, other notable researchers such as Ben Shneiderman \cite{shneiderman1990human}, Helen Nissenbaum \cite{johnson1995computers} and Batya Friedman \cite{friedman1993discerning} have also promoted similar ideas that ``systems could be designed in compliance with ethical ideals such as privacy and user autonomy" \cite{van2007ict}.

Popular technology design methodologies such as Values Sensitive Design have used values as the fundamental concept to inform ethical and responsible technology design for over two decades \cite{van2007ict}. We elaborate on the relationship of values with software development methodologies in the following sections.

\subsection{Agile Methods and SAFe}\label{sec:AgileSAFe}

 %

Agile software development appeared in the mid-1990s as a response to plan-driven approaches in software engineering \cite{williams2003agile}. Since then, it has been increasingly adopted by organisations to iteratively and incrementally develop and deliver software-intensive systems\cite{hoda2018rise}, \cite{dingsoyr2012decade},\cite{versionone13th}. 

Broadly speaking, Agile development is characterised by four core values \cite{fowler2001agile}: ``\textit{individuals and interactions over process and tools, customer collaboration over contract negotiation, working software over comprehensive documentation}, and \textit{responding to change over following a plan.}''.
Respect, openness, transparency and trust are
core Agile values.


The concept of `Values' has a different connotation in the Agile community compared to a social sciences or psychology standpoint.
As discussed in Section \ref{sec:ValueDefinition}, social sciences have a broader view on human values, one that encompasses anything that humans consider as important, without any moral imperative. 
At their core, Agile values \textit{do} share some common grounds with human values of trust, helpfulness, respect, honesty, creativity, and so on. However, the Agile community tends to use Agile values as synonyms for ``principles'' which may be very specific to one particular methodology.
Agile values encourage continuous discovery of `better ways of developing software' compared to how it is done in heavyweight methodologies. A wide range of Agile methods\cite{abrahamsson2017agile}, such as eXtreme Programming (XP)\cite{beck2000extreme}, Scrum\cite{rubin2012essential}, and Crystal methodologies\cite{cockburn2004crystal} manifest the core values and principles of the Agile Manifesto\cite{williams2003agile},\cite{dingsoyr2012decade}. While the software industry has witnessed both successful and failed Agile adoptions\cite{boehm2002get},\cite{dikert2016challenges}, Scrum, or its combination with other Agile methods, remains the most widely adopted Agile development method in the industry\cite{versionone13th}. 


The benefits of Agile methods experienced by small teams led to an increasing interest of large organisations in agility\cite{dikert2016challenges},\cite{conboy2019implementing}. Nonetheless, some characteristics of large organisations make it difficult to adopt the Agile methods designed for small, co-located teams in large organisations with several teams \cite{boehm2002get}, \cite{dikert2016challenges}. Increased coordination and communication needs at the inter-team level, between software development teams and other units within the organisation (e.g. human resources), and geographically distributed settings\cite{dikert2016challenges},\cite{paasivaara2018large} imply Agile practices need to scale beyond single team application. Several frameworks have been proposed for this purpose, such as Scaled Agile Framework (SAFe)\cite{SAFe2020}, Large-Scale Scrum (LeSS)\cite{lessFramework2020}, and Disciplined Agile Delivery (DAD)\cite{ambler2012disciplined}. These frameworks aim to scale up the existing Agile practices and ceremonies or introduce new roles, practices, and ceremonies particularly suited to the large-scale context \cite{dikert2016challenges}.


In particular, SAFe has become a very popular approach for industrial large-scale agile transformations \cite{versionone13th}. SAFe is an elaborate, comprehensive framework based on Agile and Lean principles\cite{SAFe2020},\cite{turetken2017assessing}. It promotes four core values: \textit{alignment, built-in quality, transparency, and program execution} with the aim to help organisations assess their large-scale Agile transformation and to satisfy their business goals\cite{SAFe2020}. 

\begin{figure*}[htb]
	\centering
    \includegraphics[width=1\textwidth]{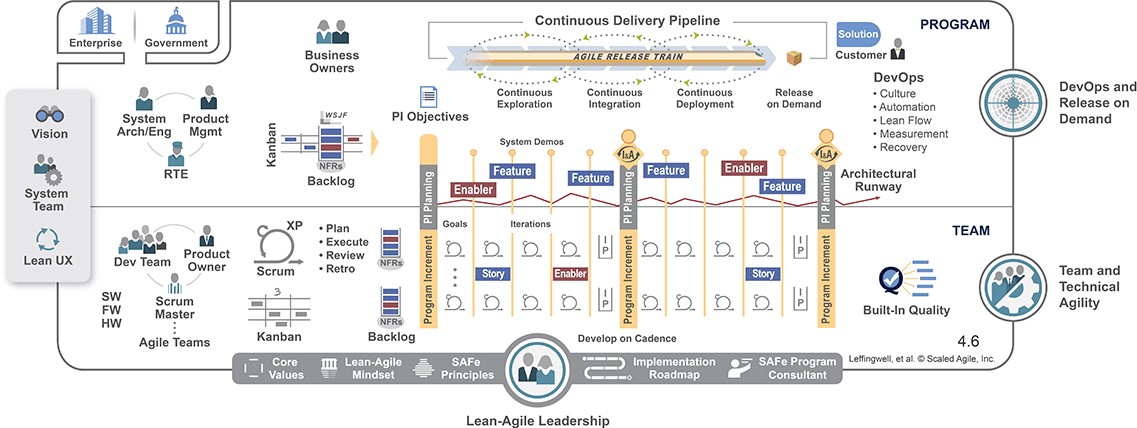}
	\vspace{-3mm}
   \caption{Essential SAFe Configuration (V4.6) (adopted from \cite{SAFe2020})}
	\label{fig:SAFe}
\end{figure*}

SAFe version 4.6 comes with four `out-of-the-box' configurations that organisations can adapt based on the scale and complexity of the systems they develop. From the basic to the most comprehensive, these configurations are (a) \textit{Essential} \textit{SAFe}, (b) \textit{Large Solution} \textit{SAFe}, (c) \textit{Portfolio}\textit{ SAFe}, and (d) \textit{Full SAFe}, respectively. The case company has adopted \textit{Essential} \textit{SAFe}; the simplest starting point to implement SAFe. As shown in Figure \ref{fig:SAFe}, the Essential configuration operates under two coordination levels namely, \textit{Team level} and \textit{Program level}.
\begin{itemize}
   \item The \textbf{\textit{team level}} includes Agile teams (i.e., each has 7 to 9 members with all necessary roles) that use Scrum or Kanban methods to collaboratively develop and iteratively release parts of a solution. Here, solution can be a service or a product released to the customer. The integration among Agile teams is facilitated through synchronised, fixed-length iterations. This level provides a wide range of \textit{build-in-quality} practices for Agile teams; for example, Agile teams can leverage user stories from XP but deliver value in Sprints adopted from Scrum\cite{SAFe2020},\cite{turetken2017assessing}. 
 \item SAFe introduces Agile Release Train (ART) at the \textbf{\textit{program level}} to organise and synchronise Agile teams. All Agile teams in an ART follow a fixed schedule to release a piece of work \cite{ebert2017scaling}.
Each release at this level is planned and executed by all relevant Agile teams (typically 5 to 12 teams) and stakeholders.
\end{itemize}

A recently published review study\cite{dikert2016challenges} found that there is little evidence-based knowledge about the usage of Agile scaling frameworks. The authors call for more empirical
studies to understand how Agile scaling frameworks are adopted in the industry and how they may be customised. Nevertheless, agile methods are popular with practitioners and therefore are a critical subject of study in software engineering.

\subsection{Human Values in Software Development Process}\label{sec:ValueSEProcess}
Human values are discussed as a part of technology design since the 1970s \cite{van2015design}.
\textit{Value Sensitive Design (VSD)} is recognised as one of the most prominent design methodologies to embed social and moral values in technology development~\cite{van2015design}. To date, VSD has mainly been applied in human-computer interaction (HCI)~\cite{Davis2015}. Despite a few isolated attempts to apply VSD in the software development life cycle (SDLC)~\cite{friedman2002value, van2015participatory}, it is yet to gain a foothold in day to day software development practice~\cite{aldewereld2015design}.
Moreover, to the best of our knowledge, there is no systematic approach in the mainstream SE practice that embeds human values into SDLC from start to finish, i.e., from requirements elicitation to software delivery. However, a limited number of human values such as security, privacy, and accessibility, has been studied in SE as software quality attributes \cite{whittle2019your}. 


A few emerging studies have attempted to incorporate human values into different phases of SDLC. For example, Thew et al. \cite{thew2018value} proposed the \textit{Value-Based Requirements Engineering (VBRE)} method to 
guide the analyst on how to elicit values, emotions, motivations from 
requirements engineering artefacts. Other techniques such as \textit{Values-First SE}~\cite{ferrario2016values} and \textit{Values Q-sort}~\cite{winter2018measuring} emphasise that values should be one of the main drivers in SE decision making. These techniques address one phase of SDLC. 
In contrast, Aldewereld and colleagues attempted to cover all phases of SDLC with their~\textit{Value-Sensitive Software Development (VSSD)} framework~\cite{aldewereld2015design}, which has not yet gained wide popularity or adoption among practitioners. Overall, these techniques are either at a conceptual level or require industrial trials before they can be part of everyday software engineering practice.

\subsection{Human Values and Agile Methods}\label{sec:ValueAgile}

Miller and Larson\cite{miller2005agile} argued that human values and ethical principles should be the backbone of evaluating Agile methods. This can help and inform developers to have more productive discussions about the strengths and weaknesses of Agile methods. They suggested two ethical analysis techniques for this purpose: \textit{utilitarian analysis} and \textit{deontological analysis}. The former helps software engineers understand and analyse the potential consequences of their actions and decisions on the stakeholders who are directly and indirectly affected by the software. The latter focuses on the intentions of software engineers and encourages them to consider human values and human needs as the leading driver in their decisions. 

Several studies have attempted to extend or augment Agile methods to put the needs 
of their customers and end-users at the centre of the software development process, particularly at the design phase\cite{brhel2015exploring},\cite{schon2017agile}. In an extensive investigation of literature, Schön et al.\cite{schon2017agile} highlighted \textit{Human-Centered Design, Design Thinking, Contextual Inquiry}, and \textit{Participatory Design} as prominent approaches used in Agile software development for this purpose. However, these methods focus more on commercial or business value for the customers and are less concerned, if at all, with human values. 

More recent studies 
have attempted to address human values in agile software development. Detweiler and Harbers\cite{detweiler2014value} proposed the idea of \textit{Value Stories Workshop} to guide Agile teams in identifying and analysing what the stakeholders of a system consider as important in their daily life (i.e., human values). Fagerholm and Pagels\cite{fagerholm2014examining} explored the values of developers working in Lean and Agile contexts and compared them with the basic human values (i.e., the Schwartz values). Their study, which is based on the perspective of 61 developers, shows that Lean and Agile values can be organised into 11 groups (e.g., \textit{the freedom to organise}, \textit{collaboration} and so on). 
Fagerholm and Pagels found that while there exists a relation between Lean and Agile values (e.g., \textit{reliance on people}) and the basic human values (e.g., \textit{benevolence}), these are two different concepts. 

In another study\cite{lawrence2012interpretation}, Lawrence and Rodriguez analysed two sources of Agile documentation (i.e., the vendor-sponsored whitepapers and IT and business magazine articles) to understand how basic values in the Lasswell value framework are interpreted and expressed by the Agile community. The Lasswell value framework includes eight values: \textit{power}, \textit{respect}, \textit{rectitude}, \textit{affection}, \textit{well-being}, \textit{wealth}, \textit{skill}, and \textit{enlightenment}. The study found that \textit{power, enlightenment, wealth}, and \textit{skill} are the most frequently expressed values among the contributors to the Agile method. However, the level of correlation of these four strong values to the values expressed in the Agile Manifesto is different. \textit{Enlightenment} and \textit{skill} are highly correlated to the Agile Manifesto, while \textit{wealth} and \textit{power} are not explicitly mapped to the values in the Agile Manifesto. It was also observed that \textit{rectitude, respect, affection}, and \textit{well-being} were less expressed. 

To the best of our knowledge, we are the first to empirically investigate the need for human values in  agile methods, in particular, in SAFe, and to map a comprehensive set of opportunities in SAFe where human values can be effectively embedded.

\section{Research Method}
\label{sec:CaseStudyDesign}

This research uses a qualitative, exploratory case study \cite{yin2014case} to investigate \textit{where} and \textit{how} to introduce human values in the Scaled Agile Framework (SAFe). Exploratory case study aims to investigate a phenomenon, in the form of a causal relationship, that has not been widely researched \cite{yin2014case}, \cite{mills2009encyclopedia}. 
We decided to employ exploratory study since they serve as a prelude to social research \cite{tellis1997introduction}, this truly fits the objectives of our study since the social construct of human values is generally an unexplored phenomenon in large scale agile development.
The exploratory nature of the study is aimed at informing software engineering practice with respect to human values; a concept which currently does not enjoy an established theory with SE \cite{perera2020},\cite{ponelis2015using}.
Furthermore, our research questions (e.g., asking \textit{how} questions) can be more appropriately answered using an exploratory approach.

Overall our qualitative approach facilitated a holistic and in-depth understanding of the software practitioners' perspective of human values derived from the direct conversations we had with the participants \cite{merriam2015qualitative}, \cite{ponelis2015using}.

We utilised Schwartz theory of values that explains the nature of and origins of human values and the underlying motivational goals to further strengthen our understanding of practitioners' values systems and hierarchies \cite{schwartz2001value}.
Because SAFe does not explicitly address human values, a study of how human values are \textit{currently being addressed in SAFe in practice} was not possible. However, the combined experience, expertise, and wisdom of SAFe practitioners can be harnessed to explore how human values \textit{can be addressed} in SAFe. This is what we did.

\subsection{The Case Company}\label{sec:CaseCompany}
Selection of a relevant case was crucial for our study. Our primary criterion for selecting a case was to identify a software company that had articulated their corporate values and was interested in including these values into their software development methodologies. This could enable us to investigate how human values translated into SE practices.

We purposefully selected an IT business unit (we refer to it as the case company) of a large public organisation who met our primary selection criteria that is, they had codified their organisational values in their `Cultural Handbook' to govern their development operations. As values are an emergent concept in the software industry, randomly choosing a case might not offer adequate insights into the phenomenon of interest. On the other hand, a purposive selection approach which we applied in this research, maximises the likelihood of collecting relevant information from the participants \cite{easterbrook2008selecting} and aids the inferential process \cite{seawright2008case}.

The selected case company has a significant SE operation and mainly does in-house software development but also utilises external contractors. It provides a wide range of IT services for the public organisation with a clear values statement. While the large public organisation was founded in the 1950s and currently has 18000 staff, their IT business unit is relatively new and consists of approximately 500 staff. We chose this case company for our study as it was willing, able and enthusiastic to support the research.

The case company was previously applying a mix of Agile and Waterfall methods for their development but has been using the Essential configuration of SAFe \cite{leffingwell2018safe} for its projects and services development for the last three years. 
Essential configuration is the simplest starting point of SAFe implementation, which \textit{``is the basic building block for all the other configurations''} \cite{leffingwell2018safe}. The company has applied a Lean-Agile Leadership Model as well as Team and Technical Agility principles as prescribed in Essential SAFe. Within SAFe, product teams at the case company use a combination of practices from Scrum, Lean, and Kanban and utilise human-centred design and design thinking.

\begin{table*}[htb]
\centering
\caption{Interviewee demographics (M: Male, F: Female, y: Year, m: Month)}
\label{tbl:demographic}
\vspace{-3mm}
\resizebox{\textwidth}{!}{%
{\renewcommand{\arraystretch}{1.5}
\begin{tabular}{llccclllccc}
\cline{1-5} \cline{7-11}
\multicolumn{1}{c}{\multirow{2}{*}{\textbf{ID}}} & \multicolumn{1}{c}{\multirow{2}{*}{\textbf{Role}}} & \multirow{2}{*}{\textbf{Gender}} & \multicolumn{2}{c}{\textbf{Experience in}} &  & \multicolumn{1}{c}{\multirow{2}{*}{\textbf{ID}}} & \multicolumn{1}{c}{\multirow{2}{*}{\textbf{Role}}} & \multirow{2}{*}{\textbf{Gender}} & 
\multicolumn{2}{c}{\textbf{Experience in}} \\
\multicolumn{1}{c}{} & \multicolumn{1}{c}{} &  & \textbf{IT} & \textbf{Case Co.} &  & \multicolumn{1}{c}{} & \multicolumn{1}{c}{} &  & \textbf{IT} & \textbf{Case Co.} \\ \cline{1-5} \cline{7-11} 
I01 & Release Train Engineer & M & 22y 9m & 1y 6m &  & I09 & Strategic Business Analyst & F & 24y & 2y 1m \\
I02 & Scrum Master & F & 22y 5m & 1y 3m &  & I10 & SAFe Agile Trainer & M & 17y 9m & 4y 3m \\
I03 & Business Transformation Manager & F & 20y & 20y &  & I11 & Product Manager & F & 20y 11m & 5y 6m \\
I04 & Digital Transformation Manager & M & 8y 10m & 2y 9m &  & I12 & Strategic Business Analyst & F & 14y 1m & 7y 11m \\
I05 & Business Analyst & M & 12y & 2y &  & I13 & Developer & M & 2y 1m & N/A \\
I06 & Change Specialist & F & 9m & N/A &  & I14 & Release Train Engineer & M & 28y 2m & 10m \\
I07 & Strategic Business Analyst & M & 19y 2m & 6y 1m &  & I15 & Delivery Lead & F & 7y 10m & 7y 10m \\
I08 & Business Change Specialist & F & 11y 9m & 3y 4m &  & I16 & Chief Information Officer & F & 18y 2m & 10y 8m \\ \cline{1-5} \cline{7-11} 
\end{tabular}%
}\quad
}
\end{table*}

\subsection{Data Collection}\label{sec:DataCollection}
 The main source of data was from the semi-structured interviews and observations. Other sources of data for this study included digital artefacts.


\subsubsection{Semi-Structured Interviews}
\label{sec:Semi-Structured-Interviews}
\textbf{Protocol:} Interviews consisted of open ended questions to understand perceptions of software practitioners regarding values. The main theme of interview enquiry was related to participants' views on areas within SAFe (e.g. roles, practices, tools or events) to introduce human values. The general pattern of the interview was to provide a brief background of our research objectives and human values, and ask about the roles and responsibilities of the interviewee in the case company. This was followed by questions relating to how human values can be introduced into the existing SAFe framework (i.e., anywhere from project inception to software delivery that may include software artefacts, development tools, practices or people). Examples of questions we asked from the participants during interviews include: 

\begin{itemize}
\item \textit{What are your personal and organisational values?}


\item In your opinion which ethical or moral values should guide software development?
\item \textit{Can you identify some of the potential intervention points within the SAFe framework where human values can be considered/introduced?} 
\item \textit{In your opinion, how can the interventions you identified be introduced in practice?}
\end{itemize}

We followed the best practice interviewing techniques recommended in \cite{patton2002qualitative} to allow adaptability to the roles and individual experiences of the interviewees but at the same time making sure that relevant areas of our research were explored. All interviews were carried out face-to-face where possible, else Skype or Zoom video calls. 
Each interview lasted between 30 and 45 minutes. Interviews were audio-recorded and used for analysis.

In line with our project's ethics approval, research participants were provided the project's explanatory statement, outlining brief but necessary information on the nature and objectives of the investigation, mutual expectations, and sample interview questions. While the concept of human values was introduced and explained in the explanatory statement, we sent additional reading material to the participants on values and Schwartz's theory prior to the interviews. We referenced the same material if the participants did not manage to read through it before the interview or required some clarification about the project or the notion of values during the interview. Prior to concluding the interviews, we asked participants for suggestions to improve our interview process (e.g., communication style, timing, the pace of the interview) and used them to improve our future interviews.  



\textbf{Participants:} Once University ethics approval was obtained, we distributed research study descriptions to potential participants by working through the company point-of-contact. Interested participants were asked to directly contact the researchers via email. To take part in the research, the participants had to be members of an active software product team.
All participants who replied and fulfilled this criterion were subsequently interviewed. In total, we interviewed 16 software practitioners with an average of 16 years of experience in the IT industry. The majority of the participants were female (9 participants, 59\%), while the rest were male. A range of software engineering practitioner roles was interviewed, including Developer, Delivery Lead, Release Train Engineers, Scrum Master, Digital Transformation Manager, Business Analysts, SAFe Agile Trainer, and the CIO (see Table \ref{tbl:demographic}).
\begin{figure}[bth]
	\centering
    \includegraphics[width=0.485\textwidth]{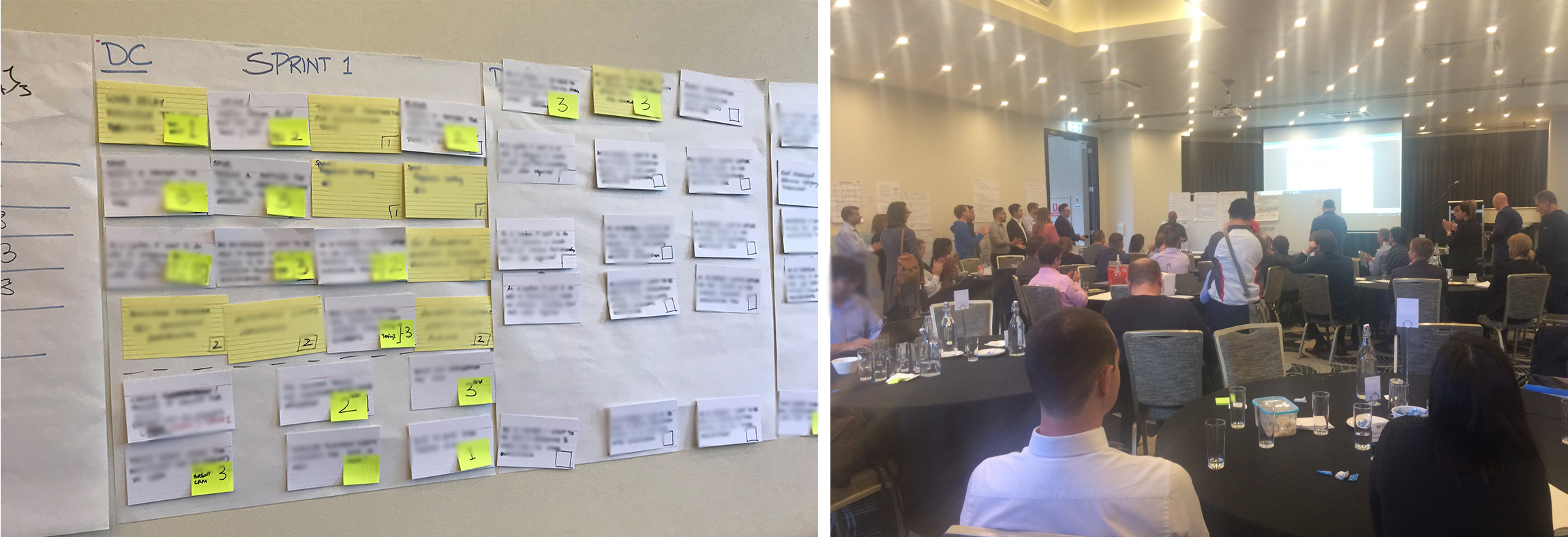}
   \caption{A snapshot observation of a PI planning meeting}
	\label{fig:observation}
\end{figure}
\begin{figure*}[htb]
	\centering
    \includegraphics[width=1\textwidth]{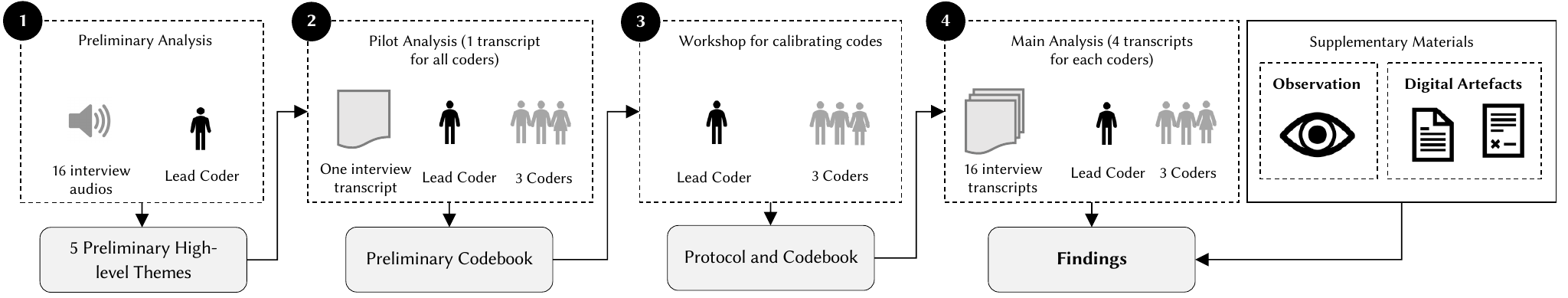}
   \caption{Data analysis process using Thematic Analysis}
	\label{fig:dataanalysis}
\end{figure*}


\subsubsection{Observations}
\label{sec:Observations}
\textbf{Protocol:} 
The case company allowed researchers to join three types of team events: daily stand-up, pre-program increment planning (PI), and PI planning. In our case company, stand-ups were used mainly to discuss development 
updates and issues as well as to plan for the next 24 hours. 
Pre-PI planning meetings were held once every three months where key members of cross-functional agile teams participated in order to align work tasks and coordinate milestones to create realistic objectives for the PI planning session. This event was followed by a major PI planning event, which is at the heart of SAFe \cite{SAFe2020}.


Our observation of the four PI planning meetings was of the longest duration (approximately 45 hours in total). Each PI planning meeting in the case company span over two working days. Figure \ref{fig:observation} shows a snapshot of one of our PI planning observations. It hosted 25 teams (i.e., each team had approximately 10 members) from the same ART (Agile Release Train) to synchronise the teams to deliver a shared vision. The activities therein included: announcements from the organisational executives, status reporting from the program leadership, planning activities (feature and user stories development for the PI), and discussions about software artefacts such as storyboards, feature plans, workflow, PI objectives, Burn Down Charts, resource capacity, and allocation. 

Our observations of team events allowed us to gain first-hand insight into the extent to which SAFe-prescribed activities were followed during these events. These observations also allowed us to collect first-hand information on whether human values were (explicitly or implicitly) discussed during these events. We observed that the conversations during these events mainly revolved around product/software functionality, teams/work coordination and dependencies, delivery goals and milestones, and project status and plans. Rather than an explicit discussion about values, occasionally other related concepts such accessibility, reliability or robustness, or the ‘purpose’ of PI were observed to be listed on storyboards and discussed. Moreover, the observations helped us understand the perceptions of the participants regarding human values, which they might have been unable to communicate through the interviews \cite{malhotra2007marketing}.

The researchers took pictures and field notes during the observation of these events and activities. 
We utilised the opportunities to attend SAFe ceremonies of our case organisation to further interact with the research participants. We asked questions to aid our understanding of the activities that took place and clarify concepts such as PI objectives, NFRs, technical debt, etc. This proved quite helpful in our analysis. The participants approached during these events included business owners, delivery leads, and current and former CEOs.

\subsubsection{Digital Artefacts}\label{sec:DigitalArtefacts}
The case company also made available additional data sources such as project plans, vision statements, user stories, change documents, and issue logs. Data from the digital artefacts were used to triangulate findings from other sources, such as the interviews and observations.

\subsection{Data Analysis}\label{sec:DataAnalysis}
Thematic Analysis (TA) was used to analyse the qualitative data collected from the interviews \cite{braun2006using}. The NVivo software was used to facilitate the coding process across the research team. As shown in Figure \ref{fig:dataanalysis}, the data analysis process was carried out in four steps. In the first step, the first author (i.e., lead coder), who conducted the interviews, listened to the audio recordings of the 16 interviews to become familiar with the interview data and extract the dominant themes. This step resulted in the construction of five preliminary high-level themes to describe the intervention points within SAFe where human values can be introduced: (a) \textbf{\textit{software/IT artefacts}}, (b) \textbf{\textit{(practitioner) roles}}, (c) \textbf{\textit{SAFe group activities/ceremonies}}, (d) \textbf{\textit{software practices}} including Lean, Scrum and Kanban practices, and (e) \textbf{\textit{tools}}. In Step 2, the lead coder and three authors of the paper (coders) conducted a pilot coding exercise on a common transcribed interview. While the coders were given the five high-level themes to guide their data analysis process, they were free to add new codes and themes. 



Step 2 was followed by a half-day workshop (Step 3), where all coders compared the similarities and differences of their coded transcript with each other and explained their rationales. A codebook was produced as an outcome of the workshop that included five themes, (a) to (e) mentioned above, and the underlying codes along with their descriptions. In the next step, each coder transcribed four interviews assigned to them and used the codebook to separately code the assigned interviews.
The lead coder was additionally responsible for overseeing and ensuring consistency across all coders. The coders held weekly meetings to discuss their work status and resolve issues encountered during coding. As the analysis progressed, the \textit{Tools} theme was dropped from the codebook because of the lack of enough evidence. \textit{Organisational culture} was added as a new high-level theme to the codebook due to repeated patterns found in the data.
Table \ref{tbl:codingexample} shows examples of open coding.
Furthermore, two types of triangulation were used in this study \cite{carter2014use}: (a) \textit{investigator triangulation}, by involving multiple researchers to bring confirmation of findings and to benefit from multiple perspectives; and (b) \textit{data source triangulation}, by collecting data from multiple sources (i.e., interviews, observations, and digital artefacts). 
Most supplementary material confirmed our findings from the interviews. Where we found any discrepancy, we consulted with participants and made necessary corrections. An earlier version of this paper was shared with the case organisation for confirmation and feedback. The participants mainly reviewed our high level findings and checked for any potential breach of the signed non-disclosure agreement. Furthermore, we shared our findings with one of the leaders in the case organisation by means of a presentation. We received positive comments and encouraging feedback regarding our research and findings.


\begin{table*}
\centering
\caption{Examples of open coding}
\label{tbl:codingexample}
\vspace{-3mm}
\resizebox{1.03\textwidth}{!}{%
{\renewcommand{\arraystretch}{1.5}
\begin{tabular}{p{5.5cm}p{3.7cm}p{4.4cm}p{2.3cm}l} 
\hline
\textbf{Interview Excerpt}                                                                                                                                       & \textbf{Code}                                                                       & \textbf{Refined Code}                                                                            & \textbf{Intermediate Theme}                  & \textbf{Theme}     \\ 
\hline
\textit{"Just add the value to the [user] story."}                                                                                                               & Adding values to user stories                                              & Modifying existing artefacts: user stories                                              & Artefact Modification/Valuefication & Artefact  \\
\textit{"You might equip 3 or 4 people to go back to their teams and try and implement the values and meeting with this mentor person so then they’ll discuss."} & Adding a new role- Values Mentor                                           & Specialised Values Roles: Mentor                                                        & Specialised values-based Role       & Role \\
\textit{"I would even bring it [values] back to testing.... actually testing against human values requirements"}                                                 & Adding values to testing and validation criteria                           & Modifying testing practices: Values Based Testing, Values Based Requirements Validation & Practice Modification/Valuefication & Practice  \\
\textit{"If the foundation and culture of the team remain consistent, people are more likely to assimilate into a founded culture"}                             & Keeping cultural foundations consistent                                    & Building consistency in the team's culture                                              & Fostering values culture            & Culture   \\
\textit{"That's not going to happen in one hit right at the end. It's got to be continuously revisited every sprint every PI"}                                   & Maintaining ongoing values communication to influence software development & (Multi-layered multi-faceted) ongoing values communication                              & Communication culture               & Culture   \\

\hline
\end{tabular}
} \quad
}
\end{table*}
\begin{table*}[]
\centering
\caption{Findings Summary: intervention points in SAFe to introduce human values in software \newline
\faLightbulbO: new ideas emerged;  \faEdit: modifying existing items; \faUsers: Culture; T: SAFe \textbf{Team} Level; P: SAFe  \textbf{Program} Level}
\label{tab:SummaryResultsNew}
\vspace{-3mm}
{\renewcommand{\arraystretch}{1.3}
\begin{tabular}{p{17cm}c}
\hline
\textbf{Intervention point} & \textbf{ Level}  \\ \hline 
\rowcolor{lightgray}\multicolumn{2}{c}{\textbf{Artefacts}} \\ \hline
\textbf{\faEdit {} User story:} state human values in the `\textit{so that}' part of a user story to capture its purpose or the bigger `why'. & T   \\
\textbf{\faEdit {} Epic/Feature:} state human values as high level requirements in the epic/feature definition. & T   \\
\textbf{\faEdit {} Persona:} state users values as higher-order goals in persona definition as potential expectations to be satisfied by the system 
.& T   \\
\textbf{\faEdit {} User Journey Map:} describe the values of users affected by system usage when they try to achieve their goals.& T   \\
\textbf{\faLightbulbO{} Checklist:} list values to be explicitly considered during development and validated through users feedback of the system.& T   \\
\textbf{\faLightbulbO{} Legislative Requirement:} incorporate regulatory guidelines 
into development processes to avoid violations of privacy/security etc.& P   \\
\textbf{\faLightbulbO{} Corporate Directive:} introduce corporate directives encouraging values consideration in software beyond security/privacy.& P   \\ \hline
\rowcolor{lightgray}\multicolumn{2}{c}{\textbf{Roles}} \\ \hline
\textbf{\faEdit {} Product Owners:} optimise business value by ensuring human values are also addressed in the system delivered.& T   \\
\textbf{\faEdit {} Business Owners:} provide inputs on what satisfies users values expectations to maximise value (creation) for stakeholders.& T   \\
\textbf{\faEdit {} Release Train Engineers:} ensure alignment between the Agile process and human values is never derailed during development.& T   \\
\textbf{\faLightbulbO{} Values Champions:} actively advocate for values inclusion and keep it alive in people's minds to ensure success.& T   \\
\textbf{\faLightbulbO{} Values Mentors:} guide development teams and encourage continuous adherence to values while developing software.& T   \\
\textbf{\faLightbulbO{} Values Translators:} communicate values concepts in a terminology understandable for the development teams.& T   \\
\textbf{\faLightbulbO{} Values Promoters:} advocate for inclusion of human values in software and help overcome barriers encountered.& T   \\
\textbf{\faLightbulbO{} Values Officers:} answer values-specific questions and ensure values related directives and guidelines are adhered to.& T   \\ \hline
\rowcolor{lightgray}\multicolumn{2}{c}{\textbf{Ceremonies}} \\ \hline
\textbf{\faEdit {} Stand-ups:} use them to assess and align the development work with human values or a `higher' propose' stand-ups should uphold.& P   \\
\textbf{\faEdit {} Inspect and Adapt:} reflect on current practices and propose new ways to embed human values in software. & P  \\
\textbf{\faEdit {} PI Planning:} align organisational vision with stakeholder values to plan software delivery accordingly. & P   \\
\textbf{\faEdit {} \textbf{Kick Offs:}} allow key stakeholders to define major human values objectives for the entire project. & P   \\
\faLightbulbO{} \textbf{Values Conversations:} prioritise human values and contemplate the implications of values breaches by software use.& T 
\\ \hline
\rowcolor{lightgray}\multicolumn{2}{c}{\textbf{Practices}} \\ \hline
\textbf{\faEdit {} Business Facing Testing:} seek user feedback beyond usability, functionality to capture users experiences of their values expectations. & T   \\ \hline
\rowcolor{lightgray}\multicolumn{2}{c}{\textbf{ Culture}} \\ \hline
\textbf{\faUsers {} Hiring/Induction:}
hire people for person-organisation 'values-fit' and foster values-driven culture through awareness and training. & P\\
\textbf{\faUsers {} Leadership:} role-model values-based thinking and action as an example behaviour for the (development) teams to follow.& P   \\
\textbf{\faUsers {} Collective Responsibility:} share responsibility to address human values throughout the development life cycle.
& P  \\ \hline
\end{tabular}%
} \quad
\end{table*}

\section{Findings}\label{sec:Findings}


Based on rigorous analysis of the data collected from the interviews, observations, and digital artefacts, our findings reveal that SAFe primarily focuses on business values and \textbf{\textit{lacks human values}} (Section \ref{sec:SAFeLack}). We present a comprehensive set of interventions to introduce human values in SAFe, which emerged from our data analysis, to introduce human values in \textbf{\textit{artefacts}} (Section \ref{sec:Artefact}), \textbf{\textit{roles}} (Section \ref{sec:Roles}), \textbf{\textit{ceremonies}} (Section \ref{sec:Ceremonies}), \textbf{\textit{practices}} (Section \ref{sec:Practice}), and \textbf{\textit{culture}} (Section \ref{sec:Culture}). In doing so, we address our overarching research question, \textbf{\textit{where and how can human values be most effectively addressed in the Scaled Agile Framework}}? 
Table \ref{tab:SummaryResultsNew} presents an overview of these intervention areas, which are detailed in the following subsections. We include several quotations from the participant interviews as a glimpse into their perceptions and ideas and as examples of the underlying empirical data collected and analysed in our study.



\subsection{SAFe lacks Human Values}\label{sec:SAFeLack}


As summarised in Section \ref{sec:AgileSAFe}, SAFe promotes four `values' \textit{alignment, built-in quality, transparency,} and \textit{program execution} for organisation-wide adoption that primarily represent technical and business concerns of software delivery. 

The Delivery Lead (I15), for example, confirmed that \textit{``SAFe is very much focussed on technical delivery, regardless of whether it says it is not. It is very much that the whole framework has been designed around getting an efficient and effective enterprise-wide delivery mechanism put in place''}. She elaborated that it does not allow human values to be sufficiently addressed in practice: \textit{``It is a framework that can be used for many purposes, but the way it is trained, the way it is described, and all of the artefacts that it asks you to produce are based on an agile framework which is all about software delivery. It doesn’t introduce the roles, and it doesn’t introduce specifically the considerations that you are after in the human values space, it just doesn’t do that''}.

Other participants perceived SAFe to be less accommodating towards Human-Centred activities. They pointed out the absence of human-centred design and design thinking from the 
framework itself and found it hard to simply `fit in'. The Release Train Engineer contended that `out of the box' SAFe framework ignores the human component and does not allow human values integration into software delivery, making it rather `\textit{soulless}':

\textit{``Some of the key pieces are missing from the soul of SAFe. It is very much focused around the practices and principles that deliver business value, which makes it a little \textbf{soulless} because it ignores the human component''} Release Train Engineer (I14).


He added that software teams' alignment to the organisational values remains unaddressed in SAFe: ``\textit{There is no mention as to how do teams organise to the organisations' values and behaviours, it is not part of any of the documentations either''} Release Train Engineer (I14).
 
While human values can potentially be introduced anywhere during software development, participants suggested that interventions would only work if they were made in the areas and in ways practitioners were already familiar with.

\textit{``The major thing is that we need to frame the conversation in something they already know and something that their framework already handles and manages''} S. Business Analyst (I09).

When categorising the suggested interventions into themes, it became apparent that most fit under the interventions made to software \textit{artefacts, roles, ceremonies, practices}, and \textit{culture}. In the following sections, we describe the interventions 
suggested for each of these themes. 

\subsection{Introducing Human Values in Artefacts}
\label{sec:Artefact}

Software artefacts help communicate the design of software or describe the process of development itself. Examples include project plans, personas, class diagrams, user stories, software code, and test cases, and the like. Participants suggested 
values-based interventions made through modifications to the \textbf{existing artefacts} in SAFe or Agile as well as by introducing some \textbf{new artefacts}. 

\subsubsection{Introducing Human Values in Existing SAFe Artefacts} \label{sec:ExistingArtefact}





\begin{figure*}[htb]
	\centering
    \includegraphics[width=1\textwidth]{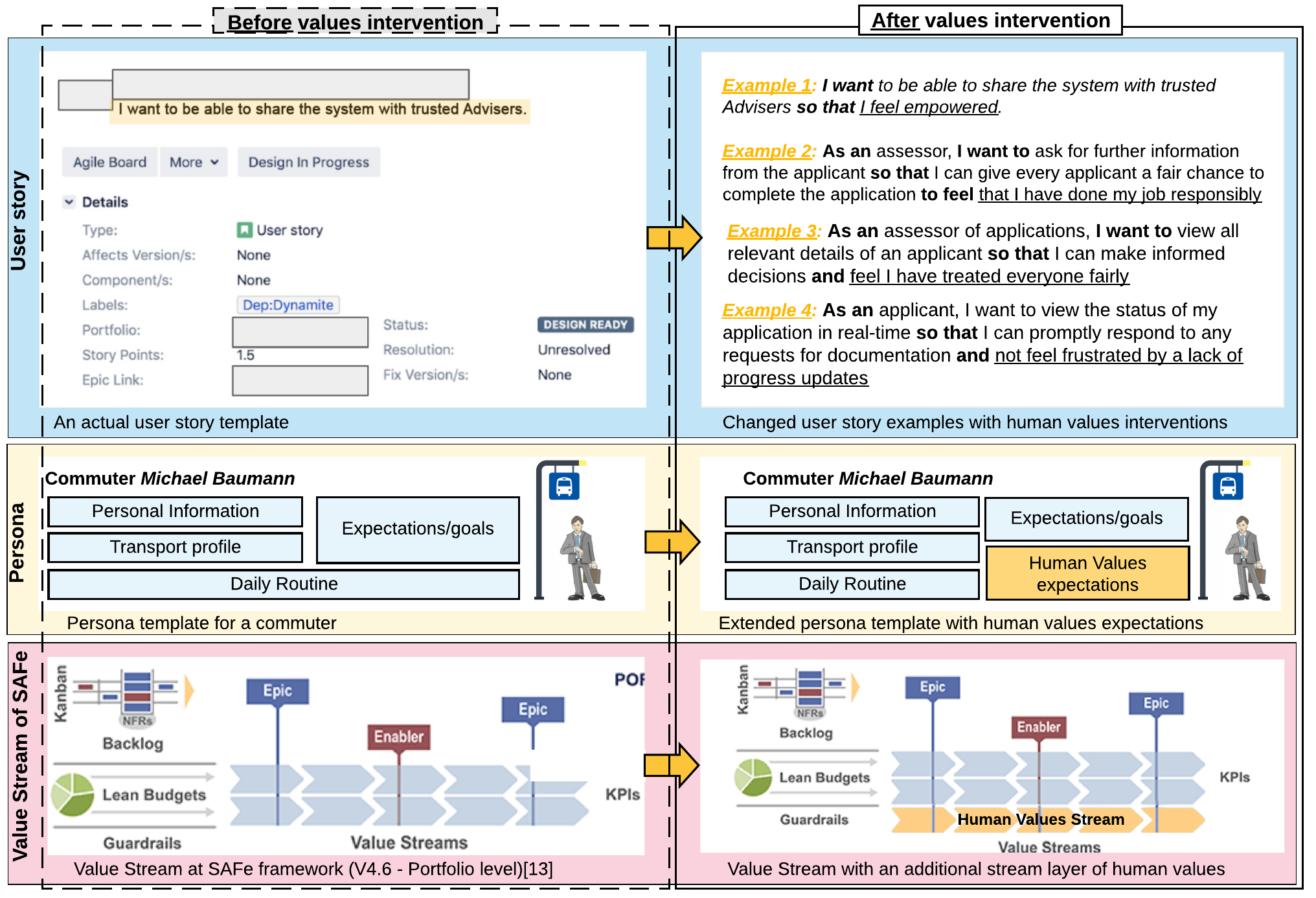}
   \caption{Examples of human values interventions in three artefacts.
   }
	\label{fig:interventionsexample}
\end{figure*}

\faEdit {} \textbf{User stories} are the most basic of Agile artefacts. In our study, user stories were identified as one of the main points of intervention by a majority of our participants (n=11, 73\%). They believed it made sense to include values into user stories as they aim to address the real needs or goals of the users (see Figure \ref{fig:interventionsexample}). They perceived goals to be the closest concept related to values, which somewhat verifies Schwartz view, from a social sciences perspective, on the goal-values relationship or proximity \cite{schwartz1987toward}.

From a more practical and pragmatic stance, the practitioners believed that for values to have a consistent representation across the board, the Delivery Lead (I15), for example, believed values need to be included at the user stories level because that is \textit{``where the rubber hits the road''}. She elaborated that from a practical perspective, including values increases the chances of their implementations because \textit{``that is the piece of work that gets done''} Delivery Lead (I15). A simplified example was provided by one of the analysts depicting how such values 'modification' to user stories can be done: \textit{``Just add the value to the story; `I want to be able to share the system with trusted advisers so that I feel \textbf{empowered}`''} Strategic Analyst (I12).

To modify a user story, she explained, first think about the ultimate user goal achieved by successfully implementing the user story and then add that as a \textit{value} to the story.  She added that while the goal of the example user story above is to deliver some functionality, the higher purpose is to enable user empowerment. Stories that deliver functionality to support or ensure users privacy and security can be similarly modified.
%

\faEdit {} \textbf{Epic} (called \textit{feature} in the case organisation) was another important artefact that could represent values as high level requirements. An epic is a work item, too large to be completed by an Agile team in one iteration and therefore has to be divided into several user stories. From an organisational perspective, the purpose or mission statements, as per Release Train Engineer (I01) can \textit{``probably translate down to the epics and capabilities in the features''.} 

From the users perspective, requirements that closely relate to values (e.g., security, privacy, usability, etc.) and are elicited during RE or assessed from user experience could be made part of epics/feature and translated into user stories.

\textit{``You can have some feature 
[or epic] statements that are talking to the human value that you want to deliver as a part of this feature''} Delivery Lead (I15).

From the development perspective, 
values need to be associated with features since they clarify the main purpose or the 'bigger why' for developing a feature. 

\textit{``Values need to be at the feature [epic] level, like we are building feature X so how does that align to the value... you should have some ability to tie it back to the value''} Developer (I13).
 
\faEdit {} \textbf{Persona} is an exemplar fictional-user that represents the characteristics and goals of a specific group of users \cite{miller2006personas}. Personas are requirements related artefacts that reflect human personality and, therefore, can portray values people may hold. Personas were believed to be a good candidate for introducing values into SAFe.
While human values were not being represented in personas in the case organisation, the participants acknowledged that their existing practice needed a change.
For example, Product Manager (I11), while reflecting over potential intervention points for values inclusion, emphasised, \textit{``I would definitely capture this as part of the persona definition''.}


Modifying personas seemed natural to include human values in software artefacts to one of the analysts. She noted that eliciting user values and reflecting them in personas should not be difficult if the analysts truly empathised with their target users to understand their real needs and concerns.
\textit{``The whole idea of personas is, you're trying to empathise with the user and you bring what user would want out of the system''} Strategic Business Analyst (I07).


\faEdit {} \textbf{User Journey Maps} are visual descriptions of the emotions a user goes through to accomplish their goal.
Making values part of journey maps was also considered important for the purpose of visualisation and teams communication. 

\textit{``...capture [values] as kind of user journey, not only the pain points or the opportunities but the emotions or the salient values [to deliver]''} Product Manager (I11).

\subsubsection{Introducing Human Values through New Artefacts} \label{sec:NewArtefact}

In addition to SAFe and Agile artefacts, participants recommended introducing other artefacts that are commonly associated with traditional disciplines and development methodologies.

\faLightbulbO{} \textbf{Checklists.} Use of checklists is common in disciplines such as civil engineering, health, and aviation \cite{clay2015back}. More recently, in machine learning based medical systems, automated implementation of checklists is considered to provide additional safety to surgeons while operating and avoid surgical errors \cite{rajkomar2019machine}. Traditionally checklists have been used for software inspection \cite{anda2002towards,brykczynski1999survey}, especially for military projects \cite{palmer1991electronic} but are not common practice in Agile. Some interviewees considered checklists should be introduced as new artefacts in SAFe to ensure values inclusion in software development.

\textit{``You could create a checklist to make sure that it [a value] is actually covered''} Delivery Lead (I15). 

\faLightbulbO{} \textbf{Legislative Requirements.}
Regulations like General Data Protection Regulation (GDPR) require organisations to safeguard public interests of privacy, data security, and autonomy. Some respondents believed such regulations are necessary to ensure human values are considered during software development.
One participant pointed out that companies fear monetary implications like \textit{``Are we going to get fined?''}, and therefore invest heavily in GDPR compliance. He added that companies are well aware that \textit{``the potential impact of not complying is greater than [profits]''} Digital Transformation Manager (I09).


\faLightbulbO{} \textbf{Corporate Directives} 
refer to written communications that govern policies and direct organisational actions. In the case company, corporate directives guided user privacy and security considerations in software development practice. Although seemingly in conflict with agile values such as team autonomy, collaboration, and freedom, enforcing compliance through executive orders was noted as a possible solution: \textit{``Compliance is hugely important. Compliance based directive would help change behaviour or would force people to incorporate values''} Business Change Specialist (I08).

Privacy directives were seen as effective in aligning software with privacy regulations and therefore were recommended as beneficial for potentially introducing them for other human values in the development process.

\subsection{Introducing Human Values through Roles} \label{sec:Roles}

Much like artefacts, participants felt the need to introduce values related roles in SAFe, arguing: \textit{``[human values] need voice. The role for that voice doesn't actually exist in the framework''} Delivery Lead (I15). 



\subsubsection{Introducing Human Values through Existing Roles} \label{sec:ExistingRoles}

\faEdit {} \textbf{Product Owner (PO)} is the core of every product development cycle. 
POs have the high level perspective of the product vision and clients' strategic needs. They liaise with the client and work closely with the development team as 1) the communicators of the product's vision or goals, 2) prioritisers of requirements or features, 3) maximisers of product's profit, 4) optimisers of  value creation in the product and development work and  5) deciders of work results. 

POs were considered suitable for keeping values as part of the vision at the front of the team's collective mind. They were seen influential enough to make values part of existing artefacts to ensure reflection and accountability: \textit{``Product owners are really, really critical to making sure that the focus on those values get retained, get written up into things like PI objectives''} Business Change Specialist (I08).

As guardians of value inclusion, they could also potentially influence and persuade the development team by pulling back everything to link them to values. A developer shared his experience from a previous company where the PO ensured the overarching messages, roughly translated to values for our purpose, were made part of their product: \textit{``Our product owner... was just like ‘these are the things we can never deviate from, `messaging must be clear', `style must conform to this', `this is how the feeling should be', like there were four key overarching messages and every decision was made in contrast to that and you were constantly pulled back to what the overarching message is rather than iterating forward and then slowly deviating overtime''} Developer (I13). 




\faEdit {} \textbf{Business Owners (BOs)} represent the stakeholders and sponsorship of the product. They guide POs on business needs. While BOs rely on the POs to provide the capacity of the team to handle work, their timely input as to what maximises the value for the stakeholders generally sets the expectation of the teams work.
For example, Business Change Specialist (I08), while reflecting on the most suitable role in SAFe or Agile, saw BOs as critical in ensuring values the product should embody: \textit{``I would expect the business owners to come in championing the vision across the team''.}

\faEdit {} \textbf{Release Train Engineer (RTE)} being the shepherds for the delivery teams, RTEs were seen as crucial to ensuring that the alignment of delivery with values is never derailed throughout the course of software development.

\textit{``I would expect RTE to be really aware of [values] as the RTE is looking at how the things are working and is checking whether they're actually on track with what they were asked to do, that they [development teams] haven't lost sight of it as they've taken on the ownership of that project or initiative''} Business Change Specialist (I08).

\subsubsection{Introducing Human Values through New Roles} \label{sec:NewRoles}
While some existing roles could be modified to focus on delivering software with considerations of human values, some new roles were still highlighted to ensure its implementation. 

\faLightbulbO{} \textbf{Values Champions.} 
Champions are individuals who informally emerge to promote an idea with conviction, persistence, and energy. They willingly risk their position and reputation to ensure the success of their cause \cite{kratzer2017social}. 
While plenty of empirical evidence exists in the organisational innovation literature of the effectiveness of this role\cite{lichtenthaler2009role}, the participants believed such roles were needed in SAFe for successful implementation of values in software.
Strategic Business Analyst (I07) noted that inclusion of values in the artefacts alone can not guarantee their inclusion in software. He explained: \textit{``Once recorded or visually captured in some SAFe artefact, values need to be 'championed' by someone who keeps them at the front of people's mind throughout the development''.} He argued further that \textit{"even when legislation and standards are available, people need to champion individual values to ensure the delivery of that value is made possible in the software''.}





\faLightbulbO{} \textbf{Values Promoters.} Promoters utilise every possible opportunity to reiterate the importance of the idea or objective they promote and overcome barriers they encounter. Unlike champions with a generalist role, contributions of promoters in projects are more specific, e.g., power promoter, expert promoter, technological gatekeeper, relationship promoter, and so on \cite{gemunden2007role}. Values need constant 'promotion' to keep them 'alive' in the team conversations, as noted by the delivery lead. She believed this increases their chances of being included in the software.

\textit{``Having one person promoting for values is better than having none...one who is charged with spruiking that, making sure that [in] every ceremony have we thought about this in key areas, PI planning and showcases, bringing it back to the users and the values.''} Delivery Lead (I15).


\faLightbulbO{} \textbf{Values Mentor.}
Agile mentors provide initial guidance and support for development teams in adopting Agile. They remove the team's misconceptions, enable teams to become confident in Agile, and encourage continued adherence to Agile practices \cite{hoda2012self}. Agility, however, does not come natural to many practitioners and software engineers, especially those trained in traditional ways of developing software. Dealing with the abstract and 'mushy' concept of values could be very daunting for software developers, according to Release Train Engineer (I14) as \textit{``their view of the world tends to be more logical more methodical.''} To work with something as abstract as values, for those who are so engineering driven, \textit{``takes a bit of time and coaching to help them understand...why are we doing what we are doing''} Release Train Engineer (I14).
A values mentor role could fulfil this need, one who 'tells the story' by drawing links between the technical work and its higher purpose (i.e., human values). The mentor does this by explaining the benefits of software use, connecting it with stakeholder values, and helping the team to become more confident in dealing with them.
Change Specialist (I06) recommended to groom mentors from the existing system by creating `\textit{values communities of practice}' that nominates Values Mentors \textit{``who meet regularly to discuss [values-related] ideas and send them back to the teams''.}


\faLightbulbO{} \textbf{Values Translators.} Translators understand the business language used by customer representatives and the technical language used and understood by the development team. Acting as a bridge between the two groups, they
help overcome the language barrier between these
groups \cite{hoda2012self}. Not having someone with the ability to understand human values from the users' perspective and explaining that in the language of developers poses significant challenges for them. Developer (I13) highlighted the obvious need for such values translator noting, 
\textit{``One key role that is missing in SAFe is [someone with] the ability to translate what the business layer is saying what the people are trying to achieve and link that [with] overarching program goal''.}

Chief Information Officer (I16) added that \textit{``they [developers] `get it' when it [value] is articulated from a business perspective''.}  
Commenting on the abilities of a values translator, Developer (I13) highlighted that it \textit{``needs to be both, strategist and delivery person''}, but acknowledged \textit{``those people are currently ‘unicorns’ they are hard to find''.}

 One participant reported that he translated \textit{trust} for developers: \textit{``[I] took the organisational statements and goals that contain corporate values and transformed them into a language that [the implementers] would understand'' } Strategic Business Analyst (I07).


\faLightbulbO{} \textbf{Values Officers.} Officer role is not part of the SAFe framework, but one such role existed in the case organisation, called the `Privacy Officer'. This individual answered questions related to privacy regulations and compliance standards to ensure regulatory directives were adhered to. The Strategic Business Analyst (I07) explained the involvement of the privacy officer role in development projects: \textit{``Part of the project governance process involves a big conversation with the universal Privacy Officer about where data is being stored what data was collected so that ends up being formally built into project management''.} Similar roles were envisaged by participants for values other than privacy and security. 

While there can be a variety of roles that can be introduced to bring about the values-based culture in the organisation, for practical reasons, however, for organisations using agile methodologies, these responsibilities could be performed by either the product owner or the business owner (or a similar role) depending on their ability, willingness, and awareness of human values.

\subsection{Introducing Human Values in Ceremonies} \label{sec:Ceremonies}

Participants generally agreed that SAFe ceremonies allow ample opportunities for technical discussions and reflections, but they could also include and benefit from values conversations to improve the alignment of software delivery with stakeholder values.
We now describe some existing and new ceremonies that could be 'valuefied', i.e., to consciously and explicitly address the values concerns of a broad range of stakeholders. 


\subsubsection{Introducing Human Values in Existing Ceremonies} \label{sec:ExistingCeremony}
\faEdit {} \textbf{Program Increment Planning.} 
In PI planning, teams a) estimate delivery targets for the upcoming increment, b) highlight their dependencies with other Agile teams, and c) come up with a set of PI Objectives (or summary of the business and technical goals) to be achieved at the end of the PI. On average, around 200 members belonging to various teams in the program participated in these meetings held every four months for two consecutive days (usually outside the organisational premises). The relatively longer duration of these ceremonies allowed the reflective as well as envisioning type of team interactions. The participants suggested that a subset of human values could be added to PI objectives.
 
\textit{``[values] really needs to be in PI planning... it gives a good leveller for everyone to go OK, we understand what the key metrics or values [are], what we are working towards''} Developer (I13). 

Participants believed the inter-team and intra-team discussions opportunities afforded by PI planning meetings could facilitate human values discussion to understand their significance for the project and how they can be weaved into the future PIs.

\faEdit {} \textbf{Inspect and Adapt Meetings.}
Agile principles emphasise reflection on the work done to improve team performance. In SAFe, the Retrospectives and  Inspect \& Adapt meetings offer opportunities for 'reflection-on-action' where agile teams reflect on the work finished (e.g., sprint or program increment) \cite{babb2014embedding}. Several recommendations to embed reflection in Agile are found in the literature \cite{babb2014embedding,cockburn2001agile}.
The reflection aspect of these meetings was appealing for practitioners as they could collectively contemplate how and where values should be introduced to have the maximum impact on software delivery. Some considered these SAFe ceremonies to be ideal for pondering over the real meaning of their work and whether that aligned with the team or organisational values.

\textit{``The places where the most reflection on how we actually work is the Retro[spective] and then the Inspect and Adapt...they are probably the two best suited ceremonies [for values consideration]''} Business Change Specialist (I08).


\faEdit {} \textbf{Daily Stand-Ups} are daily Agile team meetings to disseminate information, share work progress, flag barriers to achieving tasks, synchronise activities, and plan for the next 24 hours \cite{augustine2005agile}. Some participants believed that values could emerge in stand-ups as they allow, to some extent, an opportunity to reflect and plan ahead. Another reason could be to link the development work with a `higher purpose' and keep it fresh in the teams' collective mind. 


From the implementation aspect, it was believed that some additional questions could simply be added to the standard discussion points during stand-ups to cover human values. 

\textit{``We already have a list of the questions that we ask at stand up...so it could very much be, how is the work you are doing today upholding that value''} Strategic Business Analyst (I12).  

However, the opinion was divided on whether or not to discuss values on a daily basis. A developer pointed out that there was \textit{``too much overhead.. people will just ignore and it will be too much constantly reviewing it''} Developer (I13).




\faEdit {} \textbf{Kick Off} meeting is a widely used Agile project management tool that defines the high-level backlog for the project and the major project goals \cite{cervone2011understanding}.
Kickoffs provide an opportunity to build a consensus among key stakeholders. They were also considered to be useful for defining values objectives for the entire project, as the CTO (I16) provided an example:  
\textit{``when we do the kickoffs we can reiterate why it [values] is important... [for example] sometimes it's not about the [customer], sometimes it is about even the mental health of the professional staff member''.}
\subsubsection{Introducing Human Values in New Ceremonies} \label{sec:NewCeremony}

Some participants believed that a new ceremony was necessary to better address human values while developing software using SAFe, as discussed in the following section.

\faLightbulbO{} \textbf{Values Conversations.} Agile methods, including SAFe, provide considerable opportunities for teams for communication and conversations, e.g., daily stand-ups, planning meetings, inspect and adapt, and so on. These, however, are conventionally more focused on technical discussions regarding performance and productivity. Some participants believed a new value oriented meeting, say values conversation, should also be introduced. The purpose of such meeting would be to solely focus on discussing human values and identifying ways to embed them into software delivery.

Participants generally acknowledged that the concept of value, which is at the core of Agile methods (e.g., SAFe), is defined rather narrowly, i.e., from a business or economic perspective.  Human values are inadequately discussed often remain unrealised in the software.
\textit{``[values] .... is just not something that software delivery teams are used to having conversations around...  it is not a voice that we commonly represent in the way that the teams operate''} Delivery Lead (I15).

These thoughts were echoed by Digital Transformation Lead (I04), who elaborated further by saying \textit{``there's an opportunity for us to start talking more about not just the software but also... How did it make you feel?... and bringing in users to talk about the level of helpfulness and empowerment that they're getting''.} He suggested that values conversations need to be held with actual people:\textit{ ``these are values that you need to hear from real people''.}

\subsection{Introducing Human Values in Practices} \label{sec:Practice}


The main practice to address stakeholder values and concerns was identified as values based testing and validation. 

\subsubsection{Introducing Human Values in Existing SAFe Practices} \label{sec:ExistingPractice}

\faEdit {} \textbf{Business Facing Agile Testing.} Agile methodologies, including SAFe, utilise several types of business facing or user testing to ensure software quality and improve user acceptance. Examples include Alpha/Beta testing, prototypes, usability testing, user acceptance testing, and simulations. 
The current practice is to seek feedback on usability, functionality, and user experience. The participants believe this could very well be extended to \textit{``get some feedback to make sure that we are now considering those [values]''} RTE (I01).


While testing for functionality is necessary, testing to ascertain whether users' values expectations are met or violated by the software is even more important, as the Delivery Lead noted (I15): \textit{``I need to test whether it is functionally sound and robust but actually [also] testing against human values requirements''}.




\subsubsection{Introducing Human Values in New Practices} \label{sec:NewPractice}

No `new' practices were recommended. Participants contended that existing practices could easily be modified to discuss human values in the SAFe framework.

\subsection{Introducing Human Values in Culture} \label{sec:Culture}
Implementation of values in software was seen as the manifestation of a values-based culture introduced by organisational leadership. 

\textit{``In order for the values to come to life, it has to be entrenched in the culture''} Digital Transformation Manager (I04).  

The participants identified the following components as intervention points to strengthen organisational culture to make it values-based and fit for human values delivery in software.


\faUsers {} \textbf{Hiring and Induction.} 
Values are at the core of organisational culture and often drive actions. The participants believed a bottom-up approach to introducing a cultural transformation is needed to make it values-driven. For example, Digital Transformation Manager (I04) suggested: \textit{``In terms of delivering software and delivering solutions that bring human values to life you have to start at the most basic layer which is the people you hire how they are on-boarded into the company''.}
Similarly, having consistent values within the team was necessary, according to Developer (I13), who noted that it \textit{``speaks to the health of the team and the values of what they hold collectively''} but felt \textit{``that’s really hard to foster unless you bring people from the very very start''.}

Prioritising and hiring people with better person-organisation 'values-fit' needs to be followed by a thorough induction. Additional awareness or training may be necessary to maintain the values and purpose alignment between the organisation and the employee identified at the time of hiring. Here is what the Delivery Lead (I15) had to say regarding this: \textit{``they {employees} should be given the {values} handbook, and a much more thorough induction that sort of explains to them how the [organisation] operates and what the strategy of the [organisation] is''.}


\faUsers {} \textbf{Leadership.} 
Leaders of Agile teams set project direction, align people, obtain resources, and motivate the teams\cite{hoda2012self}. 
The participants considered that it was the responsibility of leaders to ensure values are considered in every project. \textit{``Leaders, at any capacity need to take on that role of promoting values to allow trickling down of values to team members – not a single special role designated for values''} Delivery Lead (I15). 
Since the management style in Agile is `leadership-collaboration' instead of command and control \cite{cockburn2001agile},  values-conscious behaviour 'rubs off' onto the development teams, only when it is `role modelled' by the leadership. Such behaviours do not take root if they are hurled down to the development teams as orders that have to be followed.

\faUsers {} \textbf{Collective Responsibility} in Agile methods facilitates vigilance and enhances perceived integrity through shared work ethic\cite{mchugh2011agile}. Some participants believed that the responsibility to include values in the team conversations and to embed them in software could not be placed on a single person or role. In fact, it had to be distributed and collectively shared by everyone, i.e., from the leadership down to individual team members. 

\textit{``It should be everybody, it should be all of these people in both of these layers the program and the team level, should be trying to reach this''} Business Change Specialist (I08).






\section{Discussion} 
\label{sec:Discussion}

Our study is the first to empirically investigate how human values can be introduced into software systems developed using scaled Agile methods. The previous section provided detailed findings on the recommended intervention points to introduce values in the SAFe framework by participants from our case company. The interventions relate to the inclusion of values in general. While it can be argued that in every software some `non-negotiable' values like privacy, security, social justice or fairness, should be introduced however, the participants made no such recommendations. We also believe such suggestions are beyond the scope of this study. The decision for the type of values to be included therefore rests upon the stakeholders in a given context and project.

 Our findings provide empirical evidence of the need to introduce values in Agile frameworks such as SAFe. Practitioners may be able to relate to the findings better when looking at SAFe Principles. What jumps out especially from the SAFe (4.6)  principles is its narrow view of value, defined as \textit{``the economic worth of the capability to the business and the customer"}. Business or economic value is no doubt important in commercial software development, the added emphasis that an economic rather than a human-centric view, is \textit{the} right way to go about it is hard to justify. Note the pronounced emphasis on the economic aspect in this text from the SAFe (4.6) documentation: \textit{``It [economic worth] is Principle 1 for a reason. Economics should inform and drive decisions at all levels, from Portfolio to Development Teams''}. No wonder some of the more conscientious practitioners thought that such a view/stance makes SAFe rather \textit{`soulless'}.

The recent version of SAFe (5.0) introduces customer-centricity as a `mindset' that enterprises should focus on. It promotes aligning solution delivery to end-users so that the customers enjoy a positive experience. While this is far better than the transactional or utilitarian approach discussed previously/above, the delivery of customer value in SAFe (5.0) even with newly introduced customer-centricity continues to guide the economics of the problem solving/the solution by documented statements like: \textit{``There are two primary means by which a customer derives value from products and solutions, 1) reducing their costs and 2) increasing their revenue".} This suggests that the current version of SAFe is focused on business/economic value. Hence, there is a need to adapt some existing building blocks of SAFe or introduce new ones to steer its focus on human values-based solution delivery.
\subsection{Make Human Values Actionable}
Values, admittedly are subjective. Software engineers in particular consider them as ``mushy stuff" \cite{miller2005agile,9261980}.
Since 1994, when Jim Collins and Jerry Porras published their book ``Built to Last" making the case that the best companies adhered to certain core values, many executives felt pressured to conjure up some values of their own. However, too often these value statements turn out to be rather bland or toothless; something which stands for nothing but a desire to be \textit{au courant} or politically correct\cite{9261980}. This leaves companies with an ambiguous identity, one without any rallying point for the employees. The ambiguity of values hinders the understanding necessary for their translation into business contexts and implementation in software. 

This resonated with the participants who emphasised the introduction of values oriented roles who not only genuinely believe in the values cause but have the necessary expertise or skills, for example, to help reduce the ambiguity and translate these concepts in a language that software engineers can understand and implement. This is important because defining value statements more clearly and consciously enhances their chances of implementation. Furthermore, when developed conscientiously, communicated clearly, and adhered to devoutly, values can foster a culture that sets companies apart from the competition. Therefore, instead of using vague nouns for values which often do not prompt any action, \textit{values (statements) should make sense and have to be made actionable \cite{lencioni2002make}.} 

\subsection{Evolution rather than Revolution}
Addressing human values might be seen as requiring fundamentally new software techniques and methods: a revolution in the way that software practitioners think and work \cite{8917668}. However, findings from our study indicate that human values can be incorporated in SAFe fairly easily by combining and/or adapting the existing building blocks of Agile methods: artefacts, practices, and roles with an overarching values-driven mindset.

Where existing methods fail, however, is knowing when and how to combine and/or adapt existing artefacts or techniques. We saw evidence of lost opportunities: interviewees, for example, suggest modifying user stories to ``values stories'' to capture user values, but this was not done.
This could arguably be attributed to SAFe's focus on the 'creation' of business value rather than aligning solutions delivery with human values. Evidently, the idea of value 'creation' relates more closely to the functional or economic view of value rather than a more social or ethical view of values.
With small adaptations, like spelling out human values as overarching software requirements in user stories, we can trigger conversions that can lead to some implementable functional requirements which can be `satisfied' through software. This is one example of a relatively simple adaptation of a requirements artefact can lead to a significant impact on practice, as noted by our participants in
Section \ref{sec:ExistingArtefact} (User Stories) and other evolutionary interventions noted in Figure \ref{fig:interventionsexample}. Similar adaptations are possible in existing roles and practices. 
\textit{Guidelines for adapting practices, roles, artefacts, and culture, need to be included in SAFe documentation and coaching material to facilitate values inclusion.}

\subsection{Implement a Human Values-based Culture} 
To become a values-based organisation, something else that needs to be fixed that has to do with neither the domain of development nor sound engineering practice, and that is \textit{culture}. When something is amiss with certain elements of the organisational culture it does not remain conducive to progress. This is because change in any aspect of an organisational system is inextricably linked with the cultural context, and values have a significant impact on the success and failure of change efforts \cite{danicsman2010good}. Deficiencies in organisational culture therefore, must be acknowledged and addressed, if software organisations are to move forward. 

The case organisation was on its way towards a values-based culture. This was in part evident by some of the initial steps taken by the leadership e.g. introduction of the cultural handbook of organisational values 
to guide their development practices and instigate a cultural change (noted in Section \ref{sec:CaseCompany}). It was also evident from the participants` intent manifested in their practical recommendations to `valuefy' the existing development processes and practices (i.e. to integrate values in them) as summarised in Table 3 and covered  in Section 4.2 through to Section 4.6. Among others, these steps included a) introducing multi-threaded multi-layered communication processes to understand, promote and implement human values, b) advocating shared responsibility c) the promoting the induction and on-boarding of individuals that are more aligned to the organisational values, 
and d) targeting development outcomes that are informed by the value-systems of their users and are cognisant of potential social implications of the software. These encouraging and less 'intrusive' initiatives to cultivate change are more likely to take roots even in somewhat more`technically-oriented' software teams. This in essence may be the precursor to a values-driven organisational culture. 


While the desire was present, the awareness of the importance of integrating values into the development process and understanding the implication when done wrong, e.g., for the organisation, users, and the society, varied among the participants in the case organisation. As with the other kind of organisational culture change, this is possible with the appropriate motivation, incentives, mentoring, and leadership role modelling. Since the recommended intervention points stem from our study of the SAFe framework, which is essentially based on Scrum and Kanban, recommendations highlighted in this study could easily be adapted or applied in other Agile methods.

\subsection{Cost of Interventions}
Cultural change in organisations is notoriously challenging 
and often invokes reactions of intense nature\cite{baumeister1996identity}, \cite{smollan2009organizational}. 
The recommended interventions in this study thus may bear significant monetary and emotional costs. For example, moving from essentially a profit driven towards values driven philosophy can bear a positive outcome when participants’ values are congruent with those of the organisation. This, however, may require significant investment in making the hiring, recruitment, and training processes values-oriented, e.g., hiring an ethically conscious workforce, training existing ones, and recognising and rewarding employees that exemplify behaviour aligned to the desired values (Table 3). Both monetary cost and effort overheard of such concrete initiatives may or may not be feasible for some organisations.

\subsection{Consider Human Values in the Entire Development Life Cycle} 

Even when values are considered explicitly, organisations like the one in this study tend to consider them mainly in the early phases of a project, with values discussions taking place mainly during project selection and early requirements of design stages \cite{9261980}. Occasional consideration is also given to some values like privacy and security during Alpha/Beta testing as well, but it tends to receive less attention from the team members. 
One possible reason for this is that there are more well-known techniques that apply in the early development phases, such as personas that help software engineers to understand the behaviours of users and to design efficient feedback acquisition to meet users' needs with lower functional capability during the design process. Another reason might be the development team's perceived low return on effort on downstream development activities\cite{9261980}.
Given the nature of the interventions recommended that span the whole life cycle activities and roles, our findings suggest that instead of focusing on individual development phases and development practices \textit{software methodologies need to be adapted to address values throughout the development life cycle.}
\section{Threats to Validity}\label{sec:Threats}

We discuss the potential threats to our research method and findings from the qualitative research perspective \cite{guba1981criteria}, \cite{stol2014key}.

\textbf{Transferability}.
We collected data from an IT business unit (we refer to this unit as the case company in this paper) of a single organisation with a relatively small sample size. Hence, most observations are exclusive to the case company and may not be representative of other non-IT business units in the organisation and other organisations. 
We limited the impact of this threat by interviewing practitioners with varying levels of seniority and experiences and different roles (e.g., CIO, Senior Analysts, Developers, etc.). Moreover, the research community, in particular social sciences research, values single-case studies as a potential source of significant contributions to scientific discovery \cite{flyvbjerg2006five}, \cite{kalliamvakou2017makes}, \cite{kuper2013social}. 

\textbf{Credibility}.
The approach used for data collection, developing the interview questions, and the selection of the case company and participants could introduce possible threats to the credibility of this study. We are confident that our approach of multi-source data collection and triangulation increased the plausibility of our results significantly. To mitigate the potential threats stemming from the interview questions, (1) we kept the interview questions open-ended; (2) follow-up questions were customised according to the interviewees’ responses; and (3) we allowed the interviewees to share any significant experiences they had related to integrating and/or ignoring human values in any software development methods, not limited to SAFe \cite{hove2005experiences}. Recruiting participants that could reflect unbiased opinions about integrating human values in SAFe was a challenge. The participants came from a purposively selected company that had a clear values statement. This may have triggered the participants to exaggerate the role of human values in software development methods and how they were followed in the case organisation as well as provide more socially acceptable responses (i.e., social desirability bias) \cite{furnham1986response}. We mitigated these threats to some extent by assuring the participants that their responses and personal details would not be identifiable
\cite{gousios2016work}. It was also crucial to recruit participants with the right competencies. To this end, we discussed our study's objective with the senior members of the case company before the study started, which helped them introduce the right candidates. To further alleviate this threat, we shared the relevant references on human values (e.g., Schwartz model) with the participants in advance, leading to increased preparation and engagement in discussions \cite{hove2005experiences}.

\textbf{Confirmability}. 
We minimised researcher bias by involving four coders in the analysis process (i.e., investigator triangulation) \cite{carter2014use}.
We also set up several internal discussions to review and verify the codes identified by the four coders involved in the study. We further limited this threat by sharing an early version of this paper with one of the case company representatives and seeking his feedback. Stol et al. \cite{stol2014key} argue that such data triangulation reduces the subjectivity of the researcher’s interpretation and judgement and, in turn, helps establish confirmability. 
\section{Conclusion and Future Work} \label{sec:Conclusion}

Agile methods, as the most popular software development methodologies, lack specific and specific guidelines in their current form to address human values such as accessibility, social responsibility, equality. However, their focus on people and interactions makes them amenable to adapting and incorporating human values. 
In this paper, we report findings from our qualitative case study
to reveal the potential intervention points within one of the most widely used large scale Agile methods, the Scaled Agile Framework (SAFe), that can be leveraged to address human values. 
The identified intervention points can be broadly classified into five categories: artefacts, roles, practices, ceremonies, and culture. Our study has shown that some existing artefacts (user story, epic, persona, user journey map), roles (product owner, business owner, release train engineer), ceremonies (stand-up meeting, inspect and adapt meeting, PI planning meeting, kick off meeting), and practice (business-facing testing) in SAFe can be changed to consciously and explicitly identify and address the values concerns of a broad range of stakeholders. Furthermore, we found new roles (values champion, values promoter, values translator, values offices, values mentor), artefacts (checklist, legislative requirement, corporate directive), ceremonies (values conversation) and cultural practices (hiring/induction, leadership, collective responsibility) that can be added to SAFe to support human values inclusion in software.


Although the focus of this case study was specifically on the implementation of SAFe and its capacity for taking human values into account, findings are also relevant more broadly and further research is needed to investigate Agile methodologies in general.  Key questions arising from this study are whether Agile is conducive to the incorporation of human values, or whether its core principles and the techniques developed for practice settings might inadvertently hinder the recognition of human values. 
\appendix

\textbf{Interview Questions}

\noindent \textbf{Demographic Questions}
\begin{itemize}
   \item \textit{How long have you been working in this organisation?}
   \item \textit{What project(s) are you involved in?}
   \item \textit{What is your current role and responsibilities?}
\end{itemize}
\noindent \textbf{General Questions}
\begin{itemize}
  \item  \textit{What motivated you to become a software developer?}
\end{itemize}
\noindent \textbf{Values and Society}
\begin{itemize}
\item \textit{What are your personal and organisational values?}
  \item \textit{In your opinion which ethical or moral values should guide software development?}
  \item \textit{Do your values influence or get reflected in what you develop?}
  \item \textit{Do the 5 core values listed in your department’s Cultural Handbook get reflected in the software you develop? How?}
\end{itemize}
\noindent \textbf{Personal, Organisational Values Integration}
\begin{itemize}
   \item \textit{Does your development framework (SAFe) allow you to integrate your values and that of your organisation into the software? Why and why not?}
   \item \textit{Can you identify some of the potential intervention points within the SAFe framework where human values can be considered/introduced?}
   \item \textit{In your opinion, how can the interventions you identified be introduced in practice?}
\end{itemize}
\noindent \textbf{Concluding Questions}
\begin{itemize}
\item \textit{Is there anything we should have asked but did not?}
\item \textit{Anything you want to add to what you have already said?}
\end{itemize}

\ifCLASSOPTIONcaptionsoff
  \newpage
\fi



%



\bibliography{ms.bib}

\begin{thebibliography}{100}
\providecommand{\url}[1]{#1}
\csname url@samestyle\endcsname
\providecommand{\newblock}{\relax}
\providecommand{\bibinfo}[2]{#2}
\providecommand{\BIBentrySTDinterwordspacing}{\spaceskip=0pt\relax}
\providecommand{\BIBentryALTinterwordstretchfactor}{4}
\providecommand{\BIBentryALTinterwordspacing}{\spaceskip=\fontdimen2\font plus
\BIBentryALTinterwordstretchfactor\fontdimen3\font minus
  \fontdimen4\font\relax}
\providecommand{\BIBforeignlanguage}[2]{{%
\expandafter\ifx\csname l@#1\endcsname\relax
\typeout{** WARNING: IEEEtran.bst: No hyphenation pattern has been}%
\typeout{** loaded for the language `#1'. Using the pattern for}%
\typeout{** the default language instead.}%
\else
\language=\csname l@#1\endcsname
\fi
#2}}
\providecommand{\BIBdecl}{\relax}
\BIBdecl

\bibitem{8917668}
J.~{Whittle}, M.~A. {Ferrario}, W.~{Simm}, and W.~{Hussain}, ``A case for human
  values in software engineering,'' \emph{IEEE Software}, vol.~38, no.~1, pp.
  106--113, 2021.

\bibitem{neate_2018}
\BIBentryALTinterwordspacing
R.~Neate, ``Over \$119bn wiped off facebook's market cap after growth shock,''
  Jul 2018. [Online]. Available:
  \url{https://www.theguardian.com/technology/2018/jul/26/facebook-market-cap-falls-109bn-dollars-after-growth-shock}
\BIBentrySTDinterwordspacing

\bibitem{shane_wakabayashi_2018}
\BIBentryALTinterwordspacing
S.~Shane and D.~Wakabayashi, ``The business of war: Google employees protest
  work for the pentagon,'' Apr 2018. [Online]. Available:
  \url{https://www.nytimes.com/2018/04/04/technology/google-letter-ceo-pentagon-project.html}
\BIBentrySTDinterwordspacing

\bibitem{bradshaw_howard}
S.~Bradshaw and P.~N. Howard, ``The global disinformation disorder: 2019 global
  inventory of organised social media manipulation,'' Tech. Rep., 2019.

\bibitem{galhotra2017fairness}
S.~Galhotra, Y.~Brun, and A.~Meliou, ``Fairness testing: testing software for
  discrimination,'' in \emph{Proceedings of the 2017 11th Joint Meeting on
  Foundations of Software Engineering}, 2017, pp. 498--510.

\bibitem{perera2020}
H.~Perera, W.~Hussain, J.~Whittle, A.~Nurwidyantoro, D.~Mougouei, R.~A. Shams,
  and G.~Oliver, ``A study on the prevalence of human values in software
  engineering publications, 2015 -- 2018,'' in \emph{Proceedings of the 42nd
  International Conference on Software Engineering}, 2020.

\bibitem{philbeck2018values}
T.~Philbeck, N.~Davis, and A.~M.~E. Larsen, ``Values, ethics and innovation:
  Rethinking technological development in the fourth industrial revolution,''
  Tech. Rep., 2018.

\bibitem{EuGDPR}
{European Parliament and of the Council}, ``General {D}ata {P}rotection
  {R}egulation,'' 2016, regulation (EU) 2016/679, GDPR.

\bibitem{raso2018artificial}
F.~A. Raso, H.~Hilligoss, V.~Krishnamurthy, C.~Bavitz, and L.~Kim, ``Artificial
  intelligence \& human rights: Opportunities \& risks,'' Tech. Rep., 2018.

\bibitem{sirur2018we}
S.~Sirur, J.~R. Nurse, and H.~Webb, ``Are we there yet? understanding the
  challenges faced in complying with the general data protection regulation
  ({GDPR}),'' in \emph{Proceedings of the 2nd International Workshop on
  Multimedia Privacy and Security}, 2018, pp. 88--95.

\bibitem{hoda2018rise}
R.~Hoda, N.~Salleh, and J.~Grundy, ``The rise and evolution of agile software
  development,'' \emph{IEEE software}, vol.~35, no.~5, pp. 58--63, 2018.

\bibitem{fowler2001agile}
M.~Fowler and J.~Highsmith, ``The agile manifesto,'' \emph{Software
  Development}, vol.~9, no.~8, pp. 28--35, 2001.

\bibitem{SAFe2020}
\BIBentryALTinterwordspacing
{Scaled Agile Framework (SAFe)}. [Online]. Available:
  \url{https://www.scaledagileframework.com}
\BIBentrySTDinterwordspacing

\bibitem{williams2003agile}
L.~Williams and A.~Cockburn, ``Agile software development: it’s about
  feedback and change,'' \emph{IEEE computer}, vol.~36, no.~6, pp. 39--43,
  2003.

\bibitem{alahyari2017study}
H.~Alahyari, R.~B. Svensson, and T.~Gorschek, ``A study of value in agile
  software development organizations,'' \emph{Journal of Systems and Software},
  vol. 125, pp. 271--288, 2017.

\bibitem{sutcliffe2011experience}
A.~Sutcliffe, S.~Thew, and P.~Jarvis, ``Experience with user-centred
  requirements engineering,'' \emph{Requirements Engineering}, vol.~16, no.~4,
  pp. 267--280, 2011.

\bibitem{beck2001manifesto}
K.~Beck, M.~Beedle, A.~Van~Bennekum, A.~Cockburn, W.~Cunningham, M.~Fowler,
  J.~Grenning, J.~Highsmith, A.~Hunt, R.~Jeffries \emph{et~al.}, ``Manifesto
  for agile software development,'' 2001.

\bibitem{boehm2002get}
B.~Boehm, ``Get ready for agile methods, with care,'' \emph{Computer}, vol.~35,
  no.~1, pp. 64--69, 2002.

\bibitem{thew2018value}
S.~Thew and A.~Sutcliffe, ``Value-based requirements engineering: method and
  experience,'' \emph{Requirements Engineering}, vol.~23, no.~4, pp. 443--464,
  2018.

\bibitem{van2015design}
J.~van~den Hoven, P.~E. Vermaas, and I.~van~de Poel, ``Design for values: An
  introduction,'' in \emph{Handbook of {E}thics, {V}alues, and {T}echnological
  {D}esign: {S}ources, {T}heory, {V}alues and {A}pplication {D}omains}.\hskip
  1em plus 0.5em minus 0.4em\relax Springer, 2015, pp. 1--7.

\bibitem{fagerholm2014examining}
F.~Fagerholm and M.~Pagels, ``Examining the structure of lean and agile values
  among software developers,'' in \emph{International Conference on Agile
  Software Development}, 2014, pp. 218--233.

\bibitem{obie2021first}
H.~O. Obie, W.~Hussain, X.~Xia, J.~Grundy, L.~Li, B.~Turhan, J.~Whittle, and
  M.~Shahin, ``A first look at human values-violation in app reviews,'' in
  \emph{2021 IEEE/ACM 43rd International Conference on Software Engineering:
  Software Engineering in Society (ICSE-SEIS)}.\hskip 1em plus 0.5em minus
  0.4em\relax IEEE, 2021, pp. 29--38.

\bibitem{yin2014case}
R.~K. Yin, \emph{Case study research: Design and methods (applied social
  research methods)}.\hskip 1em plus 0.5em minus 0.4em\relax Sage Publications,
  2014.

\bibitem{bouman2018measuring}
T.~Bouman, L.~Steg, and H.~A. Kiers, ``Measuring values in environmental
  research: a test of an environmental portrait value questionnaire,''
  \emph{Frontiers in psychology}, vol.~9, p. 564, 2018.

\bibitem{usher2017social}
E.~L. Usher and D.~H. Schunk, ``Social cognitive theoretical perspective of
  self-regulation,'' in \emph{Handbook of self-regulation of learning and
  performance}.\hskip 1em plus 0.5em minus 0.4em\relax Routledge, 2017, pp.
  19--35.

\bibitem{rokeach1973nature}
M.~Rokeach, \emph{The Nature of Human Values.}\hskip 1em plus 0.5em minus
  0.4em\relax Free press, 1973.

\bibitem{schwartz1992universals}
S.~H. Schwartz, ``Universals in the content and structure of values:
  Theoretical advances and empirical tests in 20 countries,'' \emph{Advances in
  Experimental Social Psychology}, vol.~25, pp. 1--65, 1992.

\bibitem{mussbacher2020there}
G.~Mussbacher, W.~Hussain, and J.~Whittle, ``Is there a need to address human
  values in domain modelling?'' in \emph{2020 IEEE Tenth International
  Model-Driven Requirements Engineering (MoDRE)}.\hskip 1em plus 0.5em minus
  0.4em\relax IEEE, 2020, pp. 73--77.

\bibitem{schunk2012social}
D.~H. Schunk, ``Social cognitive theory.'' 2012.

\bibitem{palacin2021human}
V.~Palacin, M.~A. Ferrario, G.~Hsieh, A.~Knutas, A.~Wolff, and J.~Porras,
  ``Human values and digital citizen science interactions,''
  \emph{International Journal of Human-Computer Studies}, vol. 149, p. 102605,
  2021.

\bibitem{whittle2019case}
J.~Whittle, M.~A. Ferrario, W.~Simm, and W.~Hussain, ``A case for human values
  in software engineering,'' \emph{IEEE Software}, 2019.

\bibitem{schwartz1994there}
S.~H. Schwartz, ``Are there universal aspects in the structure and contents of
  human values?'' \emph{Journal of Social Issues}, vol.~50, no.~4, pp. 19--45,
  1994.

\bibitem{9218163}
H.~{Perera}, G.~{Mussbacher}, W.~{Hussain}, R.~{Ara Shams}, A.~{Nurwidyantoro},
  and J.~{Whittle}, ``Continual human value analysis in software development: A
  goal model based approach,'' in \emph{2020 IEEE 28th International
  Requirements Engineering Conference (RE)}, 2020, pp. 192--203.

\bibitem{Winner1980}
\BIBentryALTinterwordspacing
L.~Winner, ``Do artifacts have politics?'' \emph{Daedalus}, vol. 109, no.~1,
  pp. 121--136, 1980. [Online]. Available:
  \url{http://www.jstor.org/stable/20024652}
\BIBentrySTDinterwordspacing

\bibitem{shneiderman1990human}
B.~Shneiderman, ``Human values and the future of technology: A declaration of
  empowerment,'' \emph{Acm Sigcas Computers and Society}, vol.~20, no.~3, pp.
  1--6, 1990.

\bibitem{johnson1995computers}
D.~G. Johnson and H.~Nissenbaum, ``Computers, ethics \& social values,'' 1995.

\bibitem{friedman1993discerning}
B.~Friedman and H.~Nissenbaum, ``Discerning bias in computer systems,'' in
  \emph{INTERACT'93 and CHI'93 Conference Companion on Human Factors in
  Computing Systems}, 1993, pp. 141--142.

\bibitem{van2007ict}
J.~Van~den Hoven, ``Ict and value sensitive design,'' in \emph{The information
  society: Innovation, legitimacy, ethics and democracy in honor of Professor
  Jacques Berleur SJ}.\hskip 1em plus 0.5em minus 0.4em\relax Springer, 2007,
  pp. 67--72.

\bibitem{CommonCauseHandbook}
P.~I.~R. Centre, \emph{The Common Cause Handbook - A Guide to Values and Frames
  for Campaigners, Community Organisers, Civil Servants, Fundraisers,
  Educators, Social Entrepreneurs, Activists, Funders, Politicians, and
  everyone in between}, 2011.

\bibitem{ferrario2016values}
M.~A. Ferrario, W.~Simm, S.~Forshaw, A.~Gradinar, M.~T. Smith, and I.~Smith,
  ``Values-first {SE}: research principles in practice,'' in \emph{2016
  IEEE/ACM 38th International Conference on Software Engineering Companion
  (ICSE-C)}, 2016, pp. 553--562.

\bibitem{dingsoyr2012decade}
T.~Dings{\o}yr, S.~Nerur, V.~Balijepally, and N.~B. Moe, ``A decade of agile
  methodologies: Towards explaining agile software development,'' \emph{Journal
  of Systems and Software}, vol.~85, no.~6, pp. 1213--1221, 2012, special
  Issue: Agile Development.

\bibitem{versionone13th}
{CollabNet VersionOne}, ``13th annual state of agile report,'' Tech. Rep.,
  2019.

\bibitem{abrahamsson2017agile}
P.~Abrahamsson, O.~Salo, J.~Ronkainen, and J.~Warsta, \emph{Agile Software
  Development Methods: Review and Analysis}.\hskip 1em plus 0.5em minus
  0.4em\relax VTT Publications, 2002.

\bibitem{beck2000extreme}
K.~Beck, \emph{Extreme Programming Explained: Embrace Change}.\hskip 1em plus
  0.5em minus 0.4em\relax Addison-Wesley Professional, 2000.

\bibitem{rubin2012essential}
K.~S. Rubin, \emph{Essential Scrum: A Practical Guide to the Most Popular Agile
  Process}.\hskip 1em plus 0.5em minus 0.4em\relax Addison-Wesley, 2012.

\bibitem{cockburn2004crystal}
A.~Cockburn, \emph{Crystal Clear: A Human-powered Methodology for Small
  Teams}.\hskip 1em plus 0.5em minus 0.4em\relax Pearson Education, 2004.

\bibitem{dikert2016challenges}
K.~Dikert, M.~Paasivaara, and C.~Lassenius, ``Challenges and success factors
  for large-scale agile transformations: A systematic literature review,''
  \emph{Journal of Systems and Software}, vol. 119, pp. 87--108, 2016.

\bibitem{conboy2019implementing}
K.~Conboy and N.~Carroll, ``Implementing large-scale agile frameworks:
  challenges and recommendations,'' \emph{IEEE Software}, vol.~36, no.~2, pp.
  44--50, 2019.

\bibitem{paasivaara2018large}
M.~Paasivaara, B.~Behm, C.~Lassenius, and M.~Hallikainen, ``Large-scale agile
  transformation at {E}ricsson: a case study,'' \emph{Empirical Software
  Engineering}, vol.~23, no.~5, pp. 2550--2596, 2018.

\bibitem{lessFramework2020}
\BIBentryALTinterwordspacing
C.~Larman and B.~Vodde. {LeSS framework}. [Online]. Available:
  \url{https://less.works/less/framework/index.html}
\BIBentrySTDinterwordspacing

\bibitem{ambler2012disciplined}
S.~W. Ambler and M.~Lines, \emph{Disciplined agile delivery: A practitioner's
  guide to agile software delivery in the enterprise}.\hskip 1em plus 0.5em
  minus 0.4em\relax IBM press, 2012.

\bibitem{turetken2017assessing}
O.~Turetken, I.~Stojanov, and J.~J. Trienekens, ``Assessing the adoption level
  of scaled agile development: a maturity model for {S}caled {A}gile
  {F}ramework,'' \emph{Journal of Software: Evolution and Process}, vol.~29,
  no.~6, p. e1796, 2017.

\bibitem{ebert2017scaling}
C.~Ebert and M.~Paasivaara, ``Scaling agile,'' \emph{IEEE Software}, vol.~34,
  no.~6, pp. 98--103, 2017.

\bibitem{Davis2015}
J.~Davis and L.~P. Nathan, ``Value sensitive design: Applications, adaptations,
  and critiques,'' in \emph{Handbook of Ethics, Values, and Technological
  Design: Sources, Theory, Values and Application Domains}, J.~van~den Hoven,
  P.~E. Vermaas, and I.~van~de Poel, Eds.\hskip 1em plus 0.5em minus
  0.4em\relax Springer Netherlands, 2015, pp. 11--40.

\bibitem{friedman2002value}
B.~Friedman, P.~Kahn, and A.~Borning, ``Value sensitive design: Theory and
  methods,'' University of Washington, Tech. Rep., 2002.

\bibitem{van2015participatory}
M.~Van~der Velden and C.~M{\"o}rtberg, ``Participatory design and design for
  values,'' in \emph{Handbook of Ethics, Values, and Technological Design:
  Sources, Theory, Values and Application Domains}.\hskip 1em plus 0.5em minus
  0.4em\relax Springer, 2015, pp. 1--22.

\bibitem{aldewereld2015design}
H.~Aldewereld, V.~Dignum, and Y.-h. Tan, ``Design for values in software
  development,'' in \emph{Handbook of {E}thics, {V}alues, and {T}echnological
  {D}esign: {S}ources, {T}heory, {V}alues and {A}pplication {D}omains}.\hskip
  1em plus 0.5em minus 0.4em\relax Springer, 2015, pp. 831--845.

\bibitem{whittle2019your}
J.~Whittle, ``Is your software valueless?'' \emph{IEEE Software}, vol.~36,
  no.~3, pp. 112--115, 2019.

\bibitem{winter2018measuring}
E.~Winter, S.~Forshaw, and M.~A. Ferrario, ``Measuring human values in software
  engineering,'' in \emph{Proceedings of the 12th ACM/IEEE International
  Symposium on Empirical Software Engineering and Measurement}, 2018, pp. 1--4.

\bibitem{miller2005agile}
K.~W. Miller and D.~K. Larson, ``Agile software development: human values and
  culture,'' \emph{IEEE Technology and Society Magazine}, vol.~24, no.~4, pp.
  36--42, 2005.

\bibitem{brhel2015exploring}
M.~Brhel, H.~Meth, A.~Maedche, and K.~Werder, ``Exploring principles of
  user-centered agile software development: A literature review,''
  \emph{Information and Software Technology}, vol.~61, no.~C, pp. 163--181,
  2015.

\bibitem{schon2017agile}
E.-M. Sch{\"o}n, J.~Thomaschewski, and M.~J. Escalona, ``Agile requirements
  engineering: A systematic literature review,'' \emph{Computer Standards \&
  Interfaces}, vol.~49, pp. 79--91, 2017.

\bibitem{detweiler2014value}
C.~Detweiler and M.~Harbers, ``Value stories: Putting human values into
  requirements engineering,'' in \emph{Proceedings of the 4th International
  Workshop on Creativity in Requirements Engineering (CreaRE)}, 2014, pp.
  2--11.

\bibitem{lawrence2012interpretation}
C.~Lawrence and P.~Rodriguez, ``The interpretation and legitimization of values
  in agile's organizing vision,'' in \emph{European Conference on Information
  Systems (ECIS)}, 2012.

\bibitem{mills2009encyclopedia}
A.~J. Mills, G.~Durepos, and E.~Wiebe, \emph{Encyclopedia of Case Study
  Research}.\hskip 1em plus 0.5em minus 0.4em\relax Sage Publications, 2009.

\bibitem{tellis1997introduction}
W.~Tellis, ``Introduction to case study,'' \emph{The qualitative report}, vol.
  269, 1997.

\bibitem{ponelis2015using}
S.~R. Ponelis, ``Using interpretive qualitative case studies for exploratory
  research in doctoral studies: A case of information systems research in small
  and medium enterprises,'' \emph{International Journal of Doctoral Studies},
  vol.~10, no.~1, pp. 535--550, 2015.

\bibitem{merriam2015qualitative}
S.~B. Merriam and E.~J. Tisdell, \emph{Qualitative research: A guide to design
  and implementation}.\hskip 1em plus 0.5em minus 0.4em\relax John Wiley \&
  Sons, 2015.

\bibitem{schwartz2001value}
S.~H. Schwartz and A.~Bardi, ``Value hierarchies across cultures: Taking a
  similarities perspective,'' \emph{Journal of cross-cultural Psychology},
  vol.~32, no.~3, pp. 268--290, 2001.

\bibitem{easterbrook2008selecting}
S.~Easterbrook, J.~Singer, M.~A. Storey, and D.~Damian, ``Selecting empirical
  methods for software engineering research,'' in \emph{Guide to Advanced
  Empirical Software Engineering}.\hskip 1em plus 0.5em minus 0.4em\relax
  Springer, 2008, pp. 285--311.

\bibitem{seawright2008case}
J.~Seawright and J.~Gerring, ``Case selection techniques in case study
  research: A menu of qualitative and quantitative options,'' \emph{Political
  Research Quarterly}, vol.~61, no.~2, pp. 294--308, 2008.

\bibitem{leffingwell2018safe}
D.~Leffingwell, \emph{SAFe 4.5 Reference Guide: Scaled Agile Framework for Lean
  Enterprises}.\hskip 1em plus 0.5em minus 0.4em\relax Addison-Wesley
  Professional, 2018.

\bibitem{patton2002qualitative}
M.~Q. Patton, ``Qualitative interviewing,'' in \emph{Qualitative research and
  evaluation methods}, 2002, vol.~3, no.~1, pp. 344--347.

\bibitem{malhotra2007marketing}
N.~Malhotra and D.~Birks, \emph{Marketing Research: An Applied Approach: 3rd
  European Edition}.\hskip 1em plus 0.5em minus 0.4em\relax Pearson Education,
  2007.

\bibitem{braun2006using}
V.~Braun and V.~Clarke, ``Using thematic analysis in psychology,''
  \emph{Qualitative Research in Psychology}, vol.~3, no.~2, pp. 77--101, 2006.

\bibitem{carter2014use}
N.~Carter, D.~Bryant-Lukosius, A.~DiCenso, J.~Blythe, and A.~J. Neville, ``The
  use of triangulation in qualitative research,'' \emph{Oncology Nursing
  Forum}, vol.~41, no.~5, 2014.

\bibitem{schwartz1987toward}
S.~H. Schwartz and W.~Bilsky, ``Toward a universal psychological structure of
  human values,'' \emph{Journal of Personality and Social Psychology}, vol.~53,
  no.~3, pp. 550--562, 1987.

\bibitem{miller2006personas}
R.~Miller and L.~A. Williams, ``Personas: Moving beyond role-based requirements
  engineering,'' North Carolina State University. Dept. of Computer Science,
  Tech. Rep., 2006.

\bibitem{clay2015back}
R.~Clay-Williams and L.~Colligan, ``Back to basics: checklists in aviation and
  healthcare,'' \emph{BMJ Quality \& Safety}, vol.~24, no.~7, pp. 428--431,
  2015.

\bibitem{rajkomar2019machine}
A.~Rajkomar, J.~Dean, and I.~Kohane, ``Machine learning in medicine,''
  \emph{New England Journal of Medicine}, vol. 380, no.~14, pp. 1347--1358,
  2019.

\bibitem{anda2002towards}
B.~Anda and D.~I. Sj{\o}berg, ``Towards an inspection technique for use case
  models,'' in \emph{Proceedings of the 14th International Conference on
  Software Engineering and Knowledge Engineering}, 2002, pp. 127--134.

\bibitem{brykczynski1999survey}
B.~Brykczynski, ``A survey of software inspection checklists,'' \emph{ACM
  SIGSOFT Software Engineering Notes}, vol.~24, no.~1, p.~82, 1999.

\bibitem{palmer1991electronic}
E.~Palmer and A.~Degani, ``Electronic checklists: Evaluation of two levels of
  automation,'' in \emph{Proceedings of the International Symposium on Aviation
  Psychology}.\hskip 1em plus 0.5em minus 0.4em\relax The Ohio State University
  Columbus, 1991, pp. 178--183.

\bibitem{kratzer2017social}
J.~Kratzer and I.~Michelfelder, ``The social footprint of champions and
  promoters as creative leaders in innovating and executing,'' in
  \emph{Handbook of Research on Leadership and Creativity}.\hskip 1em plus
  0.5em minus 0.4em\relax Edward Elgar Publishing, 2017, pp. 182--202.

\bibitem{lichtenthaler2009role}
U.~Lichtenthaler and H.~Ernst, ``The role of champions in the external
  commercialization of knowledge,'' \emph{Journal of Product Innovation
  Management}, vol.~26, no.~4, pp. 371--387, 2009.

\bibitem{gemunden2007role}
H.~G. Gem{\"u}nden, S.~Salomo, and K.~H{\"o}lzle, ``Role models for radical
  innovations in times of open innovation,'' \emph{Creativity and Innovation
  Management}, vol.~16, no.~4, pp. 408--421, 2007.

\bibitem{hoda2012self}
R.~Hoda, J.~Noble, and S.~Marshall, ``Self-organizing roles on agile software
  development teams,'' \emph{IEEE Transactions on Software Engineering},
  vol.~39, no.~3, pp. 422--444, 2012.

\bibitem{babb2014embedding}
J.~Babb, R.~Hoda, and J.~N{\o}rbjerg, ``Embedding reflection and learning into
  agile software development,'' \emph{IEEE software}, vol.~31, no.~4, pp.
  51--57, 2014.

\bibitem{cockburn2001agile}
A.~Cockburn and J.~Highsmith, ``Agile software development, the people
  factor,'' \emph{Computer}, vol.~34, no.~11, pp. 131--133, 2001.

\bibitem{augustine2005agile}
S.~Augustine, B.~Payne, F.~Sencindiver, and S.~Woodcock, ``Agile project
  management: steering from the edges,'' \emph{Communications of the ACM},
  vol.~48, no.~12, pp. 85--89, 2005.

\bibitem{cervone2011understanding}
H.~F. Cervone, ``Understanding agile project management methods using scrum,''
  \emph{OCLC Systems \& Services: International digital library perspectives},
  2011.

\bibitem{mchugh2011agile}
O.~McHugh, K.~Conboy, and M.~Lang, ``Agile practices: The impact on trust in
  software project teams,'' \emph{IEEE Software}, vol.~29, no.~3, pp. 71--76,
  2012.

\bibitem{9261980}
W.~{Hussain}, H.~{Perera}, J.~{Whittle}, A.~{Nurwidyantoro}, R.~{Hoda}, R.~A.
  {Shams}, and G.~{Oliver}, ``Human values in software engineering: Contrasting
  case studies of practice,'' \emph{IEEE Transactions on Software Engineering},
  pp. 1--1, 2020.

\bibitem{lencioni2002make}
P.~M. Lencioni, ``Make your values mean something,'' \emph{Harvard Business
  Review}, vol.~80, no.~7, pp. 113--117, 2002.

\bibitem{danicsman2010good}
A.~Dan{\i}{\c{s}}man, ``Good intentions and failed implementations:
  Understanding culture-based resistance to organizational change,''
  \emph{European Journal of Work and Organizational Psychology}, vol.~19,
  no.~2, pp. 200--220, 2010.

\bibitem{baumeister1996identity}
R.~F. Baumeister and M.~Muraven, ``Identity as adaptation to social, cultural,
  and historical context,'' \emph{Journal of Adolescence}, vol.~19, no.~5, pp.
  405--416, 1996.

\bibitem{smollan2009organizational}
R.~K. Smollan and J.~G. Sayers, ``Organizational culture, change and emotions:
  A qualitative study,'' \emph{Journal of Change Management}, vol.~9, no.~4,
  pp. 435--457, 2009.

\bibitem{guba1981criteria}
E.~G. Guba, ``Criteria for assessing the trustworthiness of naturalistic
  inquiries,'' \emph{ECTJ}, vol.~29, no.~2, p.~75, 1981.

\bibitem{stol2014key}
K.-J. Stol, P.~Avgeriou, M.~A. Babar, Y.~Lucas, and B.~Fitzgerald, ``Key
  factors for adopting inner source,'' \emph{ACM Transactions on Software
  Engineering and Methodology}, vol.~23, no.~2, pp. 1--35, 2014.

\bibitem{flyvbjerg2006five}
B.~Flyvbjerg, ``Five misunderstandings about case-study research,''
  \emph{Qualitative Inquiry}, vol.~12, no.~2, pp. 219--245, 2006.

\bibitem{kalliamvakou2017makes}
E.~Kalliamvakou, C.~Bird, T.~Zimmermann, A.~Begel, R.~DeLine, and D.~M. German,
  ``What makes a great manager of software engineers?'' \emph{IEEE Transactions
  on Software Engineering}, vol.~45, no.~1, pp. 87--106, 2017.

\bibitem{kuper2013social}
A.~Kuper, \emph{The Social Science Encyclopedia}.\hskip 1em plus 0.5em minus
  0.4em\relax Routledge, 2013.

\bibitem{hove2005experiences}
S.~E. Hove and B.~Anda, ``Experiences from conducting semi-structured
  interviews in empirical software engineering research,'' in \emph{11th IEEE
  International Software Metrics Symposium (METRICS'05)}, 2005, pp. 10--23.

\bibitem{furnham1986response}
A.~Furnham, ``Response bias, social desirability and dissimulation,''
  \emph{Personality and Individual Differences}, vol.~7, no.~3, pp. 385--400,
  1986.

\bibitem{gousios2016work}
G.~Gousios, M.-A. Storey, and A.~Bacchelli, ``Work practices and challenges in
  pull-based development: the contributor's perspective,'' in \emph{2016
  IEEE/ACM 38th International Conference on Software Engineering (ICSE)}, 2016,
  pp. 285--296.

\end{thebibliography}
\bibliographystyle{IEEEtran}

%








\end{document}